\documentclass[aps,prd,twocolumn,nofootinbib,groupedaddress,floatfix,amsfonts]{revtex4}
\pdfoutput=1
\newcommand{\bo}{\raise0mm\hbox{$\Box$}}
\usepackage{graphicx}
\begin{document}

\title{Gravitational Waves From the End of Inflation: Computational Strategies}

\author{Richard Easther}
\email[]{richard.easther@yale.edu}
\author{John T. Giblin, Jr}
\email[]{john.giblin@yale.edu}
\affiliation{Dept. of Physics, Yale University, New Haven CT 06520 }
\author{Eugene A. Lim}
\email[]{eugene.a.lim@gmail.com}
\affiliation{Department of Physics and ISCAP, Columbia University, New York NY 10027} 
\date{\today}

\begin{abstract}

Parametric resonance or preheating is a plausible mechanism
for bringing about the transition between the inflationary phase and a hot, radiation dominated universe. This epoch results in the rapid production of heavy particles far from thermal equilibrium and could source a significant stochastic background of  gravitational radiation.  Here, we present a numerical algorithm for computing the contemporary power spectrum of gravity waves generated in this post-inflationary phase transition for a large class of scalar-field driven inflationary models.   We explicitly calculate this spectrum for both quartic and quadratic models of chaotic inflation, and low-scale hybrid models.  In particular, we consider hybrid models with an ``inverted'' potential. These models have a very short and intense period of resonance which is qualitatively different from previous examples studied in this context, but we find that they lead to a similar spectrum of gravitational radiation.   

\end{abstract}


\maketitle

\section{Introduction}
 
The universe has been transparent to gravitational radiation since  its energy density dropped  below the Planck scale.  Since they are unhindered by
optical opacity,  gravitational waves from the early universe can constitute
a stochastic {\it Gravitational Wave Background}, or GWB, that complements the familiar Cosmic Microwave Background (CMB).     
Like any other form of radiation, the number density of gravitons drops as the cube of the scale factor, $a$, while their energy is redshifted by a further factor of $a$, leading to $\rho_{gw} \sim 1/a^4$.   If a remnant background of gravitational radiation is laid down near the Planck scale, this would occur long before the end (or even the onset) of inflation. Consequently,  it would now be radically diluted, and thus entirely unobservable.  

 On the other side of the ledger, a number of processes in the early universe can generate gravitational waves,  yielding  a GWB which is potentially  detectable in the present epoch.  Unlike the CMB,  the GWB is strongly model dependent, and there is no consensus as to its likely form.     Since the existence of the CMB is a generic prediction of hot big bang cosmology, the detection of black body microwave radiation by Penzias and Wilson \cite{Penzias:1965wn} played a key role in establishing the big bang as the dominant cosmological paradigm. By contrast, the absence of clear expectations for the GWB implies that  a stochastic gravitational wave background which is not associated with ``standard'' astrophysical sources will provide a sensitive tool for discriminating {\it between} different models of the early universe.

   The existence of gravitational radiation is inferred from studies of binary pulsar systems, whose orbits decay at a rate entirely consistent with the energy they are expected to emit as gravitational radiation \cite{Will:2001mx}.  
A similar inference would be drawn if a B-mode polarization contribution was seen in the foreground-subtracted CMB sky.  However, in neither  case  is one  directly observing the distortions of space caused by  gravitational waves propagating through the universe.  This is the province of ``direct detection'' experiments such as  the currently operating {\it Laser
Interferometer Gravitational Wave 
Observatory} (LIGO), future space-based
observatories, like {\it Laser
Interferometer Space Antenna} (LISA) and {\it Big Bang Observer} (BBO), or detectors which are sensitive to the deformations of solid objects or cavities. The frequency range of these experiments is largely determined by their physical size.\footnote{LIGO is an exception here, as it operates in a Fabry-Perot mode, so its characteristic wavelength is actually some $10^2$ times larger than the physical size of its interferometers.}   Finally, proposals such as the Parkes Pulsar Timing Array represent a hybrid appoach, using an array of pulsars as a ``mesh'' of sensitive clocks which effectively measure changes in their distance  induced by a gravitational wave background.  This ``instrument'' is  sensitive to gravitational waves in the nano-Hertz range,  or wavelenghts of ${\cal O}( 100)$ light years \cite{Manchester:2007mx}.      

The best known candidate source for the GWB is the quantum mechanical fluctuations of spacetime during inflation, which provide a near  scale-invariant spectrum of gravitational waves in the present epoch. The {\em existence\/} and near scale-invariance of this spectrum is a key prediction of inflation.  However, the {\em  amplitude\/} of this spectrum is proportional to the inflationary energy scale, which is a very poorly constrained parameter,  as there is no consensus or  ``standard model'' of inflation.  For convenience, this amplitude  is expressed via the tensor:scalar  ratio, or $r$ -- the denominator is known to be $\sim 10^{-5}$ from measurements of the CMB and large scale structure.  Current data for the CMB B-mode   suggests that $r\lesssim .3$, which already  rules out  well-known models of inflation \cite{Spergel:2006hy,Peiris:2006sj}.      There are two contending theoretical arguments as to   the most likely value of $r$. Insisting on an inflationary potential which has a stable minimum with a vanishing potential energy (so as not to contribute a cosmological constant) and a simple algebraic form suggests that $r \gtrsim 0.001$ \cite{Boyle:2005ug}.  In practice, this limit depends very much on one's definition of simplicity.\footnote{The analysis of \cite{Boyle:2005ug} concentrates upon potentials which are no more than quartic in the inflaton $\phi$.  Models with a low scale are typically of the form $V \sim V_0 - \lambda \phi^4$ with any quadratic term strongly suppressed \cite{Easther:2006qu}.  In this case, one needs a $\phi^6$ term in order to construct a stable minimum. Since $V(\phi)$ is the {\em effective\/} potential it can contain higher order terms and/or  logarithms. This allows one to explicitly construct low-scale single field models without including non-renormalizable terms in the action, although one might object that loop contributions will only be dominant in models where the tree-level terms have been carefully tuned. However, high scale models of inflation will be severely tested by the next generation of CMB polarization experiments, and this question will ultimately be decided by experiment.}
   Conversely, attempts to build inflationary models inside string theory typically yield $r \lesssim 10^{-10}$, which can be traced to the need to avoid a trans-Planckian vev for the inflaton \cite{Lyth:1996im,Easther:2006qu,McAllister:2007bg}.  In this case the inflationary GWB is unobservable by  any proposed experiment.   Consequently,  the failure of a BBO style experiment to detect a near scale-invariant background would rule out many popular models of inflation, but would not undermine the overall paradigm.  

In this paper, we consider the generation of gravitational waves at the {\em end\/} of inflation \cite{Khlebnikov:1997di,GarciaBellido:1998gm,Easther:2006gt}.  Ironically, in models with a low inflationary scale, it turns out that this signal -- if it exists -- can naturally fall into the frequency range probed by BBO, and other future experiments including LISA and an upgraded LIGO.  By contrast, the conventional inflationary spectrum is effectively invisible in these scenarios. Consequently, this signal extends the range of models which experiments like BBO would be able to test. Moreover, since this signal depends on the details of the mechanism by which inflation ends, it provides a window into an otherwise unobservable epoch in the early universe.

In most models of inflation, the energy of the universe is dominated by the
potential energy of a homogenous and isotropic scalar field, $\phi$.  As
inflation ends, this energy must somehow be used to generate elementary particles and reheat the universe.   The specific mechanism by which reheating occurs is strongly model dependent.  
  Originally, the creation of other particle species  was thought to occur  slowly, since the inflaton field is necessarily weakly coupled to other fields.  Although the bottom of this potential has a
vanishing energy, the  field(s) oscillate
homogeneously, with their motion damped only by the expansion of the
universe.  Parametric resonance provides an efficient mechanism for extracting energy from these oscillations, and converting it into excitations of other fundamental fields \cite{Traschen:1990sw}.    This occurs even when the tree-level couplings between the inflaton and other fields are very weak, and the amplitudes of the momentum modes are approximately described by Mathieu-like functions.

In the classical picture, the momentum modes are simply the Fourier components of the corresponding field -- and if these amplitudes are growing exponentially, it follows that the fields and their associated energy densities will become increasingly inhomogeneous during the resonance phase.  This leads to the emission of gravitational radiation, and this process forms the topic of our investigations here.   The first connection between these two issues was made by  Khlebnikov and Tkachev \cite{Khlebnikov:1997di}, who predicted a gravitational wave signal whose peak amplitude was approximately $\Omega_{gw}\approx 10^{-10}$ for quartic, $\lambda\phi^4$ inflationary models, at (present day) frequencies of around 1 GHz.  This amplitude is substantial, but as  the required strain sensitivity needed to detect a signal of fixed amplitude scales as the cube of the frequency,  its detection presents an enormous technological challenge. 

In early 2006, Easther and Lim revived this topic, and argued that the amplitude of any preheating signal could be essentially independent of the inflationary scale, while its (present day) frequency was proportional to the energy scale of inflation  \cite{Easther:2006gt}. Thus, while a GUT scale model would be peaked near GHz scales, resonance following inflation at $10^9$ GeV would lead to a signal near the LIGO band, while lowers scales would overlap with the proposed range of BBO.  In  \cite{Easther:2006gt}, this proposal was investigated in a limited way by considering models where the inflationary scale differed by a factor of $\sqrt{10}$ using a fairly rudimentary numerical algorithm, derived from that employed in \cite{Khlebnikov:1997di}. This algorithm had a number of limitations, in that it did not allow the gravitational wave signal to be computed at arbitrary times, as it considers the power generated in a series of four dimensional spacetime ``boxes'', and was based on a formula for the radiated power in gravitational waves which is only strictly valid in a non-expanding universe. In the discussion below, we denote this the ``box'' algorithm.

In \cite{Easther:2006vd}, the present authors introduced a new algorithm which directly evolved the tensor component of the metric perturbation, and explicitly confirmed that a preheating signal could be visible with  future versions of LIGO or BBO when the inflationary scale is low enough, and that its amplitude was essentially independent of the inflationary scale.  This algorithm directly solves for the evolution of
the momentum-space metric perturbations sourced by the transverse-traceless
part of the stress-energy tensor, and we refer to it as the  ``spectral method.''  We assume a Friedman-Robertson-Walker
background metric with perturbations in synchronous gauge,
\begin{equation}
ds^2 = dt^2 - a^2(t) \left[\delta_{ij} + h_{ij}\right]dx^idx^j.
\label{perturbation}
\end{equation}
where $h_{ij}$ is   {\it
  Transverse-Traceless} or,
\begin{equation}
h^i_i = 0, \qquad  h^i_{j,i} =0. \label{ttcondition}
\end{equation}
The analysis here assumes that the universe has only scalar field components, but generalizes to arbitrary $T_{\mu \nu}$ -- including other situations with substantial inhomogeneity, such as phase transitions, bubble collisions or turbulence.  In numerical analysis, ``spectral methods'' refer to algorithms that decompose the quantity of interest into a set of basis functions, and solve for their coefficients to describe the system as it evolves. In our analysis, we actually evolve the fields in position space in an expanding background, using   {\sc LatticeEasy} \cite{Felder:2000hq}.    We then use this solution to source evolution of  the metric perturbations, so the overall approach is essentially a hybrid scheme.

Two further numerical algorithms have been developed to compute this signal -- a welcome development, given the intricacy of the relevant calculations.  In \cite{GarciaBellido:2007dg}, Garcia-Bellido and Figueroa restated the scaling argument of  \cite{Easther:2006gt} and computed the gravitational waves sourced by bubble-collisions at a variety of scales.   The amplitude of the signal seen in  \cite{GarciaBellido:2007dg} is not independent of the inflationary scale, while its numerical algorithm  was subsequently described  in more detail in collaboration with Sastre  \cite{GarciaBellido:2007af}.  As in our spectral approach, the nonlinear evolution of the scalar fields is used to compute the source for the $h_{ij}$. However, \cite{GarciaBellido:2007af}  evolve the perturbation in position space using a staggered leapfrog scheme, and we will refer to this algorithm as the leapfrog method.  The implementation described in  \cite{GarciaBellido:2007af} imposes the transverse-traceless constraints at discrete times during the simulation, rather than at the source level as we do. 

The final algorithm is that of Dufaux, Bergman,  Felder,  Kofman and Uzan \cite{Dufaux:2007pt}, who develop a Green's function for the tensor portion of $h_{ij}$, which are evolved in Fourier space, as in our spectral method. We refer to this as the ``Green's function'' algorithm, and it is constructed in an expanding background. The one explicit approximation in this algorithm is that it assumes the  modes which are well inside the current Hubble horizon, or $k \gg aH$, where $k$ is the comoving wavenumber, $a$ is the scale factor, and  $H$ is the Hubble parameter\footnote{ In principle, one could construct a Green's function without making this approximation, and in most cases of interest the relevant modes do satisfy   $k \gg aH$.}

 The one case which has been publicly examined using all four algorithms is  resonance following $\lambda \phi^4$ inflation. The box method (as implemented in \cite{Easther:2006gt}) yields spectra with a significantly higher amplitude than those found with the spectral or Green's function methods, but that the latter methods agree very closely with each other.  There is a distinction between methods which directly incorporate the expansion of spacetime, and we will see that the universe does grow significantly as resonance continues in the $\lambda \phi^4$ model -- and expansion tends to dilute the gravitational wave background.  Moreover,  care has to be used when  computing the source term for the gravitational waves, as we will be taking the differences of derivatives computed on a spatial lattice, which requires the subtraction of terms of similar magnitude from one another, a well known scenario for inducing numerical error.  The Green's function and spectral methods  yield very similar results for the same models, the algorithms were developed and coded independently, and are conceptually distinct. We  thus have considerable confidence that the gravitational wave background from preheating is being accurately evaluated.\footnote{The leapfrog code produces a spectrum  for the $\lambda \phi^4$ background that is qualitatively similar to that of the Green's function and spectral methods. }
 
In addition to the present approach, the direct production of metric perturbations via parametric resonance has been considered \cite{Bassett:1997gb,Finelli:1998bu,Parry:1998pn,Easther:1999ws,Bassett:1998wg,Finelli:2001db}, but this mechanism is different from that considered here, and is less likely to produce a detectable signal.  Further,  Felder and Kofman have considered the ``fragmentation'' of the inflaton following reheating, which is closely related to the source of the gravitational waves discussed here  \cite{Felder:2006cc}.

The structure of this paper is as follows. We begin by reviewing  preheating, and following this we describe the numerical computation of the transverse-traceless component of $T_{\mu \nu}$ for a scalar field dominated universe, and the subsequent computation of the gravitational wave background,  illustrating our approach with $\lambda \phi^4$  models.   We then consider models driven by a quadratic inflaton potentials,  a low scale hybrid model, and a massless quartic model with negative
coupling. The last model has a very different resonance structure from the other scenarios that have been examined in this context, and we again find a substantial gravitational wave spectrum. Finally, we summarize the numerical issues we have identified, and discuss future prospects for work in this area. 

\section{Parametric Resonance and Preheating}

\begin{figure}
\resizebox{8.4cm}{!}{\includegraphics{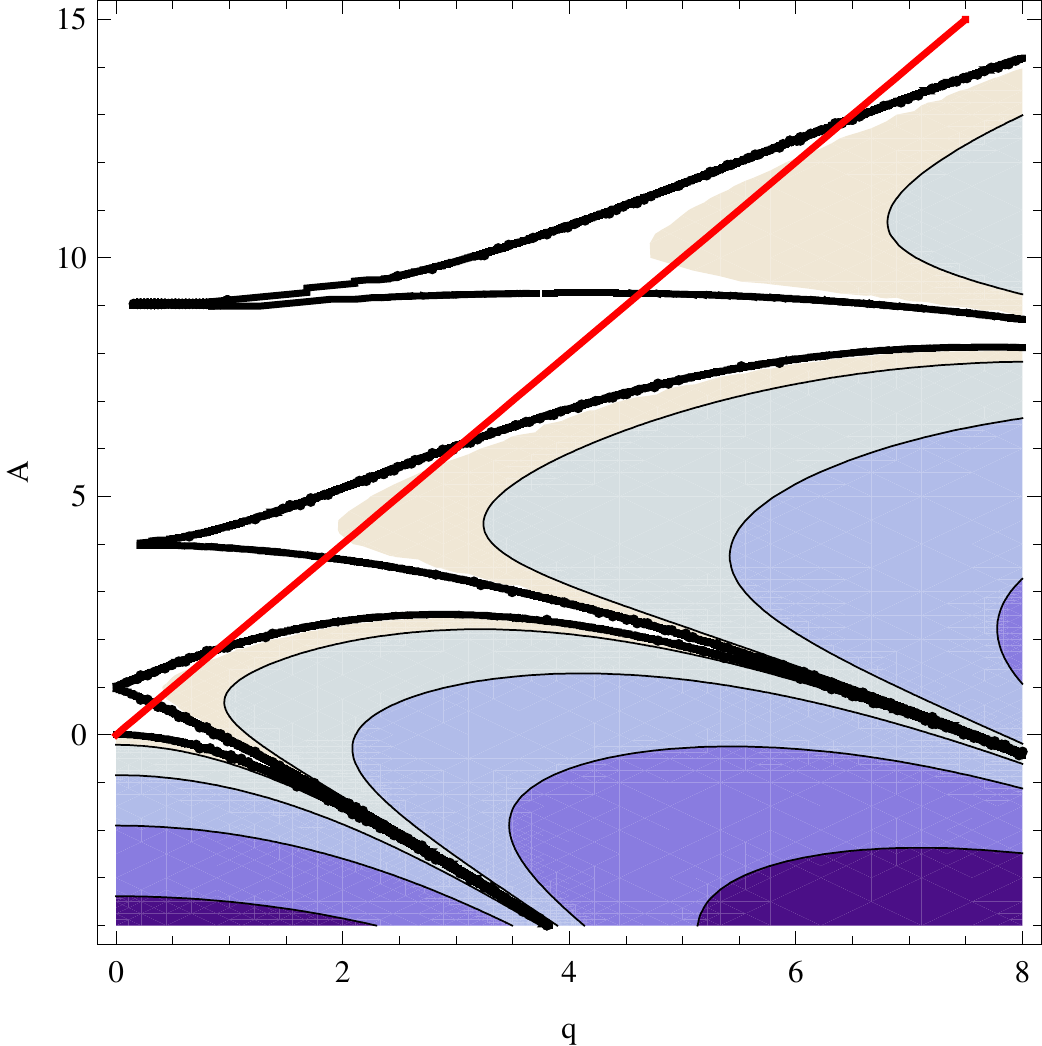}}
\caption{The Mathieu Stability/Instability chart.  The imaginary part of the Mathieu critical exponent is plotted, with darker colors corresponding to a larger imaginary component.  Outside the heavy black lines the exponent is real-valued, and the corresponding solutions are strictly oscillatory.  The diagonal line corresponds to $A_q = 2 q$.\label{stabilitychart}}
\end{figure}

We now review parametric resonance of scalar fields \cite{Kofman:1994rk,Kofman:1997yn,Greene:1997fu,Garcia-Bellido:1997wm,Greene:1997ge,Bassett:2005xm}.  Consider a toy model comprised of one {\it classical} inflaton field $\phi$ and
one (possibly massive) scalar field $\chi$
\begin{equation}
\mathcal{L} = \frac{1}{2}\partial_\mu \phi\partial^\mu \phi + \frac{1}{2}
\partial_\mu \chi \partial^\mu \chi - V(\phi,\chi),
\end{equation}
where
\begin{equation}
V (\phi,\chi) =  V(\phi) + \frac{1}{2}m_\chi^2\chi^2 + \frac{1}{2}g^2\phi^2\chi^2.
\end{equation}
In this simple picture $\phi$ is the inflaton, whereas $\chi$ is a generic ``matter'' field. 
It is not necessary to specify $V(\phi)$ at this point; however, the specific
form of $\phi-\chi$ coupling term is important.  We have additionally assumed
that there are no other $\chi$ self interactions, i.e. $\lambda_\chi \chi^4$.
Therefore $\chi$ obeys  
\begin{equation}
\ddot{\chi} +3H\dot{\chi}-\frac{1}{a^2}\nabla^2\chi + m_\chi^2\chi + g^2\phi^2\chi = 0.
\end{equation}
Adopting the Fourier transform convention
\begin{equation}
\tilde{f}(k,t) = \int d^3 x f(x,t) e^{2\pi i k\cdot x},
\end{equation}
and using $f_k$ as a shorthand for $\tilde{f}(k,t)$ we expand $\chi$ in terms of its momentum modes,
\begin{equation}
\label{eomkmode}
\ddot{\chi}_k + 3H \dot{\chi}_k+\left(\frac{k^2}{a^2}+m_\chi^2 +g^2\phi^2\right)\chi_k = 0.
\end{equation}
We can reduce this to a known analytic form by  a)  ignoring the expansion of the universe, so $H=0$ and $k/a \rightarrow k$ is simply the physical momentum, and
b) setting $m_\chi = 0$.  Typically, the end of inflation is marked by
oscillation of the inflation field about the minimum of its potential, and we will assume that this is described by a single quadratic term 
\begin{equation}
V(\phi) = \frac{1}{2}m_\phi^2 \phi^2. 
\end{equation}
With this potential  $\phi$ undergoes damped simple harmonic motion, so that 
\begin{equation}
\phi (t) = \Phi \sin(m_\phi t),
\end{equation}
where $\Phi$ is a time-dependent amplitude, which varies slowly over a single cycle.  Once we ignore the expansion of the universe, the friction term drops out of the equations of motion and $\Phi$ is strictly constant.   
Substituting into the mode equations and changing variables to 
\begin{equation}
q = \frac{g^2\Phi^2}{4m^2}, \,\,A_k = \frac{k^2}{m^2}+\frac{g^2\Phi^2}{2m^2} = \frac{k^2}{m^2}+2q, \,\, 
z=mt,
\label{mathieuparameters}
\end{equation}
turns (\ref{eomkmode}) into a Mathieu Equation \cite{bateman},
\begin{equation}
\chi_k^{\prime\prime} + (A_k-2q\cos(2z))\chi_k = 0.
\label{mathieueqn}
\end{equation}
where primes denote differentiation with respect to $z$.  The solutions to
a Mathieu equation are either stable (oscillatory) or unstable
(exponential), depending on the relative values  of $q$ and $A_q$.  Schematically, every solution to Mathieu's
equation has two parts,
\begin{equation}
\chi_k \propto f(z) e^{\pm i\mu z}
\end{equation}
where $f(z)$ is periodic. When the {\it Mathieu characteristic exponent} $\mu$ has an
imaginary part the solution has an exponentially growing mode.  Figure \ref{stabilitychart} shows the values of $\Im(\mu)$ as a function of $A_q$ and $q$.  We see that (\ref{mathieuparameters}) requires $A_q \ge 2q$, so we are interested in the parameter values that lie to the left of the diagonal line in Figure~\ref{stabilitychart}. This ensures that the we find discrete resonance bands, which grow weaker as $k$ grows larger. 

This process it not unique to $m^2\phi^2$ models. For instance, consider
\begin{equation}
\label{quartic1}
V(\phi) = \frac{1}{4}\lambda \phi^4.
\end{equation}
In this case, the oscillations after the end of
inflation are not sinusoidal; rather, they obey elliptic cosines 
\cite{Greene:1997fu},
\begin{equation}
\phi \propto cn\left(x-x_0,\frac{1}{\sqrt{2}}\right).
\end{equation}
where $x$ is a dimensionless conformal time variable defined by
\begin{equation}
x = \left(\frac{6\lambda m_p^2}{\pi}\right)^{1/4}\sqrt{t}.
\end{equation}
In terms of these new variables, the mode equations for $\chi$ are \cite{Greene:1997fu}
\begin{equation}
\chi^{\prime\prime}_k + \left(\frac{k^2}{\lambda a \phi_0} + \frac{g^2}{\lambda}cn^2\left(x,\frac{1}{\sqrt{2}}\right)\right)\chi_k = 0,
\end{equation}
where $\phi_0$ is the initial amplitude of the $\phi$ field.  This is a Lam\'e
equation \cite{bateman} which has both exponential and oscillatory solutions, and can be related to the Mathieu equation by writing
\begin{equation}
cn\left(x,\frac{1}{\sqrt{2}}\right) =
\frac{2\pi\sqrt{2}}{T} \sum_{n=1}^\infty
\frac{e^{-\pi(n-1/2)}}{1+e^{-\pi(n-1/2)}}\cos\frac{2\pi(2n-1)x}{T}.
\end{equation}
In this expression, $T$ is the period of the oscillation.  The $n=1$
term has an amplitude of $0.9550$ and dominates this series (see
\cite{Greene:1997fu} for full analysis).  If one neglects the remaining terms,
the Mathieu equation is recovered.

Generically, any periodic motion can be expanded in a sum of coherently
oscillating terms.  Each term in this expansion will have a corresponding
Mathieu equation that will lead to exponential excitation of particular
momentum modes.  However, the details of  this expansion can vary significantly, and hence different models have very different resonant behavior. When we restore the expansion of the universe and allow mass terms to be non-zero along with any nonlinear interactions, the physics grows significantly more complicated as modes move in and out of resonance \cite{Kofman:1994rk, Kofman:1997yn, Greene:1997fu}.   Further, this analysis neglects the evolution of $\phi$; like the $\chi$ field, the modes $\phi_k$ will be sourced by the $g^2\phi^2\chi^2$ coupling term.  As these modes grow,  $\phi$  becomes inhomogeneous and the coherent oscillations will break down, due to backreaction from the created $\chi$ particles. Ultimately,  the only way to follow the full evolution is via numerical simulation.

Standard treatments of parametric resonance describe the resonant pumping of momentum modes as particle creation.  However, if we simply view $\phi$ and $\chi$ as classical Klein-Gordon fields, then the amplification of their Fourier modes makes the universe increasingly inhomogeneous. The inhomogeneity in the fields leads to an inhomogeneous, time dependent energy density -- which necessarily leads to the emission of gravitational radiation.\footnote{To be pedantic, this requires a non-zero quadrupole moment, but this is generically excited by an arbitrary perturbation.} 

\section{Generation of Gravitational Power}
\label{gravpow}

\subsection{The Power Spectrum}

The stress-energy tensor associated with gravitational radiation is given by \cite{Misner:1974qy}, 
\begin{equation}
T_{\mu\nu} = \frac{1}{32\pi G} \left<h_{ij,\mu}h^{ij}_{\,\,\,,\nu}\right>,
\end{equation}
and is specific to the {\it transverse-traceless} part of the metric
perturbation.  The brackets here denote a spatial average and must be taken over a large enough volume in order to capture the contribution from long wavelength modes.  In this work, we simply integrate over the full ``grid'' on which our numerical solutions are defined.  The associated energy energy
density is just the $00$ component,
\begin{equation} 
\rho_{gw} = \frac{1}{32\pi G} \left<h_{ij,0}h^{ij}_{,0}\right> = \sum_{i,j}
\frac{1}{32 \pi G} \left<h^2_{ij,0}\right>. \label{gwdensity}
\end{equation}
Recalling Parseval's theorem we can manipulate the spatial average to find
\begin{equation}
\left<f^2(x)\right> = \frac{1}{L^3} \int d^3k |F(k)|^2,
\end{equation}
where $L$ is the length of one side of the lattice.  Thus the energy density can be expressed as an integral in momentum space,
\begin{eqnarray}
\rho_{gw} &=& \sum_{i,j} \frac{1}{L^3} \frac{1}{32\pi G} \int d^3k |h^2_{ij,0}|^2.\\&=&
\sum_{i,j}\frac{m^2_{pl}}{32 \pi } \frac{4\pi}{L^3} \int d\ln k \, k^3 |h_{ij,0}(k)|^2 
\end{eqnarray}
or more usefully
\begin{equation}
\frac{d \rho}{d \ln k} = \sum_{i,j} \frac{m^2_{pl}}{8L^3}  k^3 |h_{ij,0}(k)|^2.
\end{equation}
Lastly, this is
\begin{equation}
\label{omega}
\frac{d\Omega_{gw}}{d\ln k} = \frac{1}{\rho_{crit}}\frac{d \rho}{d \ln k} =
\frac{\pi k^3}{3H^2 L^2}\sum_{i,j}|h_{ij,0}(k)|^2.
\end{equation}
which we then insert into equation (20) of
\cite{Easther:2006gt}\footnote{Please note a typographical error appears in \cite{Easther:2006gt}} to find,
\begin{equation} 
\Omega_{gw} h^2 = \Omega_r h^2  \frac{d\Omega_{gw}(a_e)}{d\ln k}
\left(\frac{g_0}{g_*}\right)^{1/3}
\end{equation}
where $a_e$ is evaluated at the end of the simulation and $g_0/g_*$ is the ratio of number degrees of freedom today to the number degrees of freedom at matter/radiation equality and $\Omega_r$ is the present day radiation energy density. In this analysis we take $g_0/g_*=1/100$. 

\subsection{Evolution of the Perturbation}
\label{evofpert}

The goal our calculation will be to evaluate  (\ref{gwdensity}).  By construction,  the perturbation  is free of scalar and vector
components, so it represents metric deformations due only to gravitational radiation.  The perturbed Einstein's equations are
\begin{equation}
\bar{G}_{\mu\nu}(t) + \delta G_{\mu\nu}(x,t) = 8\pi G\left[
\bar{T}_{\mu\nu}(t) + \delta T_{\mu\nu}(x,t)\right]
\end{equation}
where the background $\bar{G}_{\mu\nu}$ and $\bar{T}_{\mu\nu}$ obey Einstein's
equations for the unperturbed metric, 
\begin{equation}
\label{background}
\bar{G}_{\mu\nu}(t) = {8\pi G} \bar{T}_{\mu\nu}(t),
\end{equation}
and hence,
\begin{equation}
\label{deltaeq}
\delta G_{\mu\nu}(x,t) = 8\pi G \delta T_{\mu\nu}(x,t).
\end{equation}

\begin{figure}[htbp]
\resizebox{8.4cm}{!}{\includegraphics{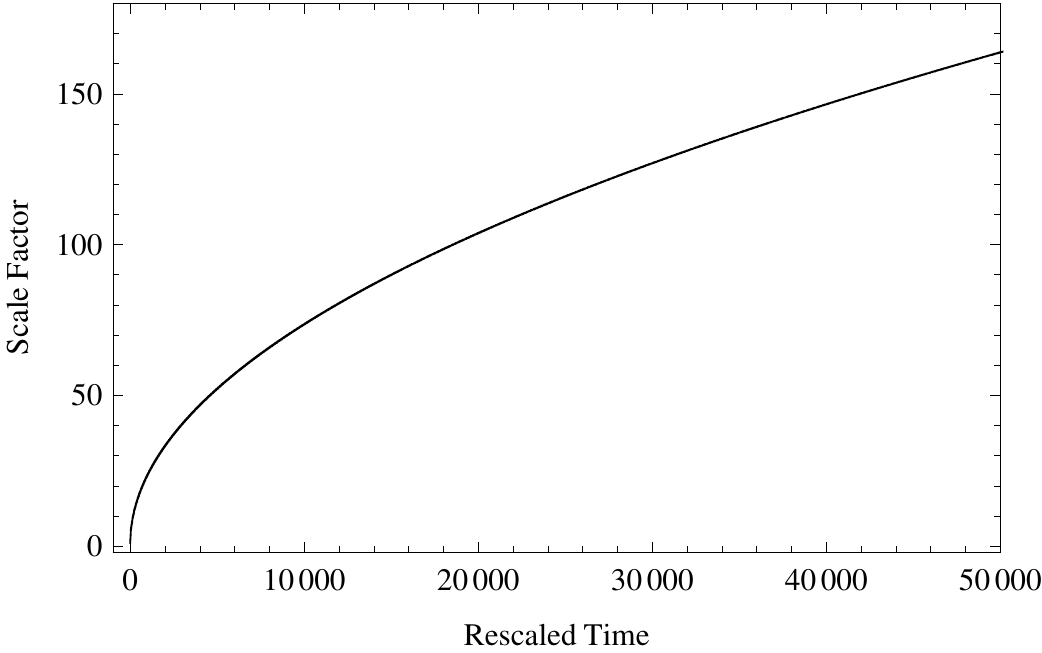}}
\resizebox{8.4cm}{!}{\includegraphics{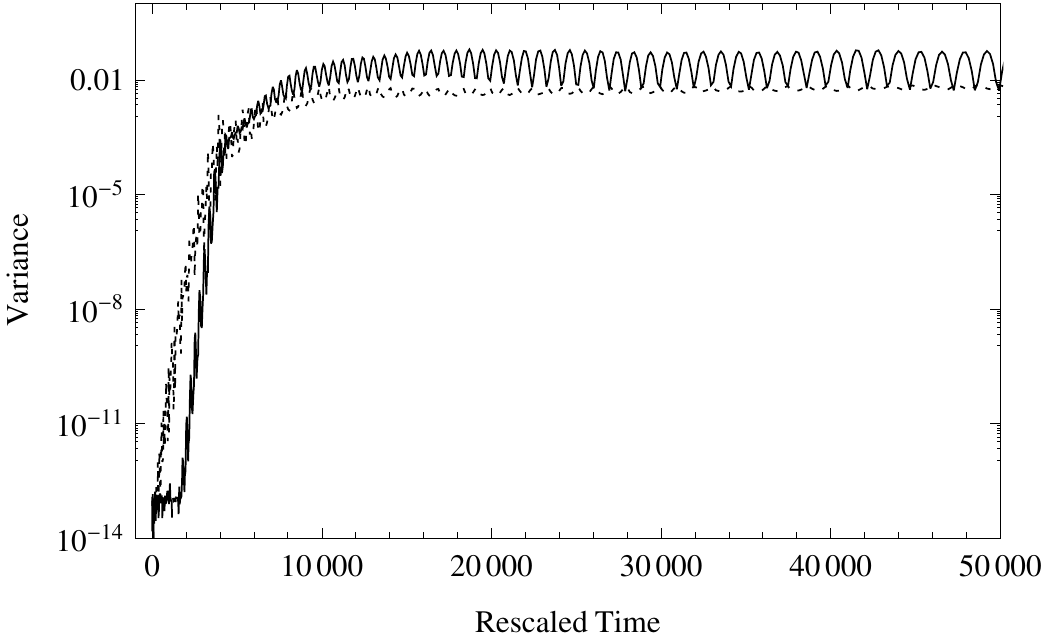}}
\resizebox{8.4cm}{!}{\includegraphics{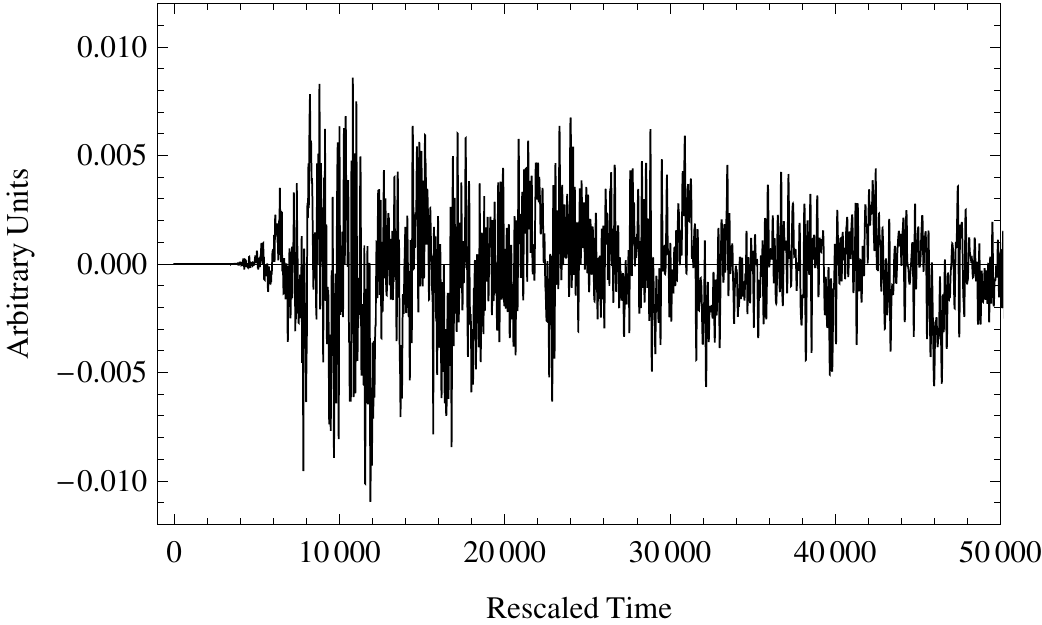}}
\resizebox{8.4cm}{!}{\includegraphics{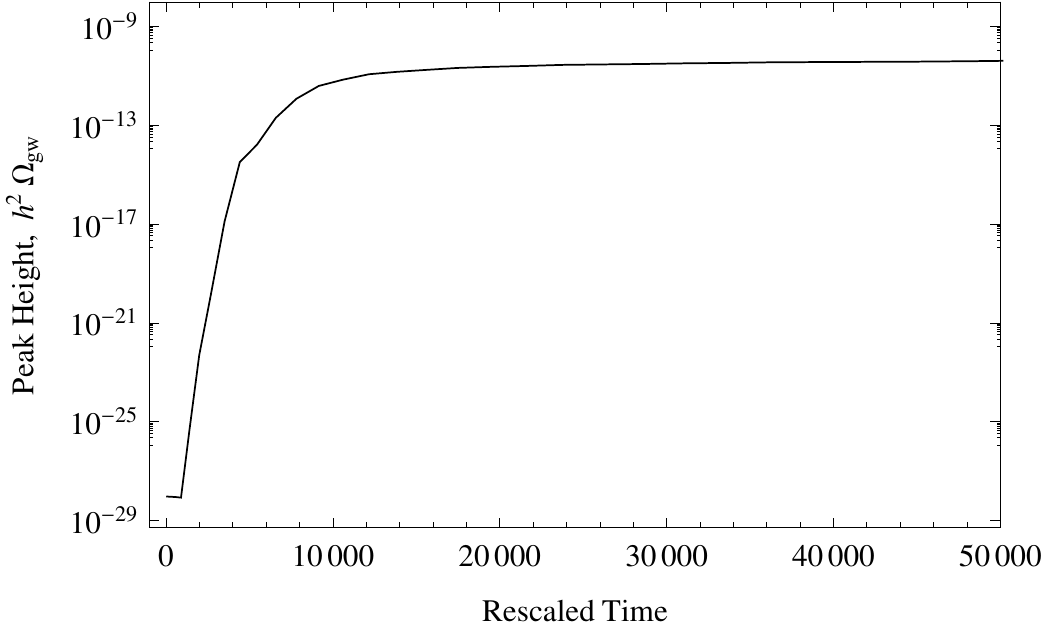}}
\caption{From top to bottom: The scale factor, the variances of $\phi$ (solid) and $\chi$  (dotted), $S_{11}^{TT}$ for a mode corresponding to $|k|\approx 1.4 \times 10^8 \,{\rm Hz}$ today, and the maximum height of the gravitational wave spectrum for this mode as a function of time for $\lambda \phi^4$ inflation with $\lambda = 10^{-14}$ and $g^2/\lambda=120$.\label{source}}
\end{figure}

One then obtains the equations of motion for the $h_{ij}$,
\begin{equation}
\label{eom}
\ddot{h}_{ij} - 2\left(\frac{\dot{a}^2}{a^2} +2\frac{\ddot{a}}{a}\right)h_{ij}
+ 3\frac{\dot{a}}{a}\dot{h}_{ij} -
\frac{1}{a}\nabla^2h_{ij} = \frac{16 \pi G}{a^2}\delta S^{TT}_{ij},
\end{equation}
where 
\begin{equation}
S_{ij} = \delta T_{ij}(x,t) - \frac{\delta_{ij}}{3} \delta T_k^k,
\end{equation} 
and $S^{TT}_{ij}$ is the transverse-traceless part of $S_{ij}$.  This can be
extracted from $S_{ij}$ -- as it can from any rank-$2$ tensor -- by projecting onto
the transverse plane and subtracting its trace \cite{Misner:1974qy}.
Explicitly, 
\begin{equation}
S^{TT}_{ij} = P_{ik}S_{kl}P_{lj} - \frac{1}{2}P_{ij}\left(P_{lm}S_{lm}\right)
\end{equation}
where the projection operator is defined by
\begin{equation}
P_{ij} = \delta_{ij} -\frac{k_ik_j}{k^2}.
\end{equation} 
This procedure is implicitly required by the form of (\ref{eom}), since one uses the transverse-traceless condition imposed on $h_{ij}$ in order to derive this from a general metric perturbation.  Consequently, we must perform the same decomposition on our source in order to consistently evolve the metric perturbations.  

  The $h_{ij}$ obey (modified) wave equations, and in principle one could evolve  $\bo h$ on the spatial lattice using the same numerical scheme employed to track the fields' evolution. In practice we found much better numerical stability when we wrote the $h_{ij}$ in terms of their Fourier transforms, and solved the resulting ordinary differential equations for $\tilde{h}_{ij}$,
\begin{equation}
\label{fourein}
\bar{G}_{\mu\nu}(\vec{k},t) = 8\pi G  
\bar{T}_{\mu\nu}(\vec{k},t) .
\end{equation}
The $k=0$ mode is our homogenous background, for which the corresponding component of  $h_{ij}$ necessarily vanishes. This is effectively a spectral algorithm for $h_{ij}$ and we use a 4th order Runge-Kutta integrator to evolve the modes.   At each point in space, we calculate $S_{ij}(\vec{x})$ explicitly from 
\begin{equation}
T_{\mu\nu} = \partial_\mu \phi_k \partial_\nu \phi_k -
g_{\mu\nu}\left[\frac{1}{2} \partial_\alpha \phi_k \partial^\alpha
  \phi_k- V(\phi_i)\right],
\end{equation}
and the full $T_{\mu\nu}$ is obtained by summing the above expression over all scalar fields.  The nonvanishing components of $S_{ij}$ are
\begin{equation}
S_{ij} = \partial_i \phi_k \partial_j \phi_k - \frac{2}{3}\delta_{ij}\left[\partial_m\phi_k\partial^m\phi_k\right].
\end{equation}
We Fourier-transform this and construct $S^{TT}_{ij}(\vec{k})$, the source term for (\ref{eom}).    

The field evolution was computed using LATTICEEASY \cite{Felder:2000hq}, which uses a {\it staggered leapfrog} integrator. The fields obey the Klein-Gordon equation in an expanding background,
\begin{equation}
\label{eomfields}
\ddot{\phi}_i + 3\frac{\dot{a}}{a}\dot{\phi}_i
 -\frac{1}{a^2}\nabla^2\phi_i +\frac{\partial V(\phi)}{\partial \phi_i} =0 \, ,
\end{equation}
where the subscript $i$ labels the fields. 
During resonance the gradient terms play a vital role, but  in their absence we recover the familiar inflationary equation of motion. As usual, $a$ is obtained from the Friedmann equations,
\begin{equation}
\ddot{a} + 2\frac{\dot{a}^2}{a}-\frac{8\pi}{a}\left(\frac{1}{3} \left|\nabla
    \phi_i \right|^2 + a V\right) =0.
\end{equation}
Consequently, these simulations are performed in an expanding, rigid spacetime and we have ignored any backreaction from metric perturbations onto the field evolution. Numerical values for $a(t)$ are normalized to unity at the beginning of our simulations, which start at the end of inflation. 

\begin{figure}[tbp]
\resizebox{8.4cm}{!}{\includegraphics{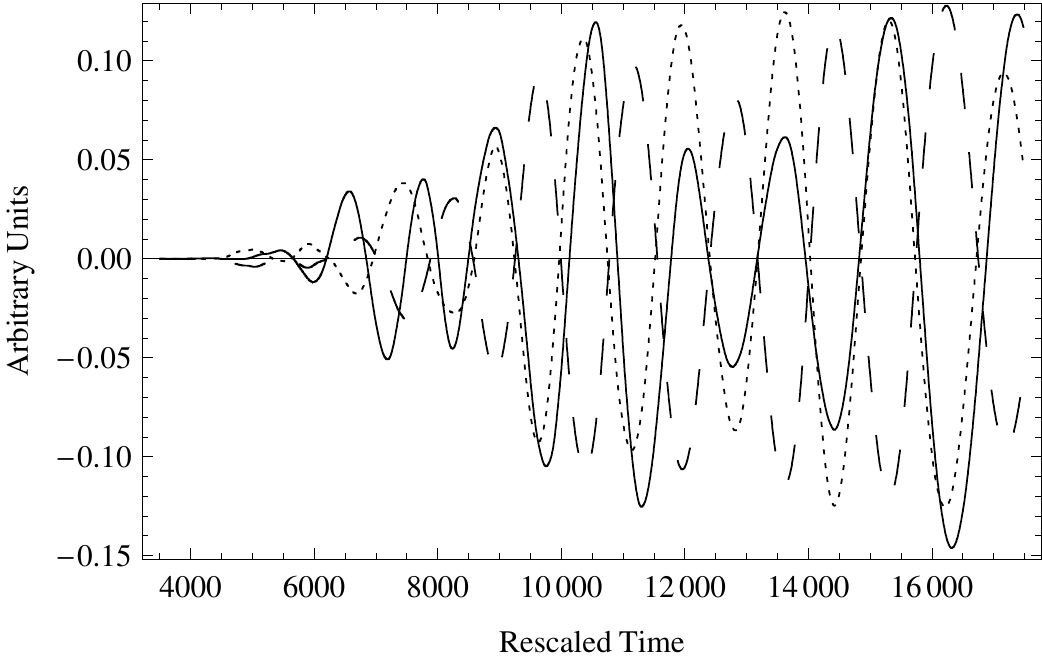}}
\resizebox{8.4cm}{!}{\includegraphics{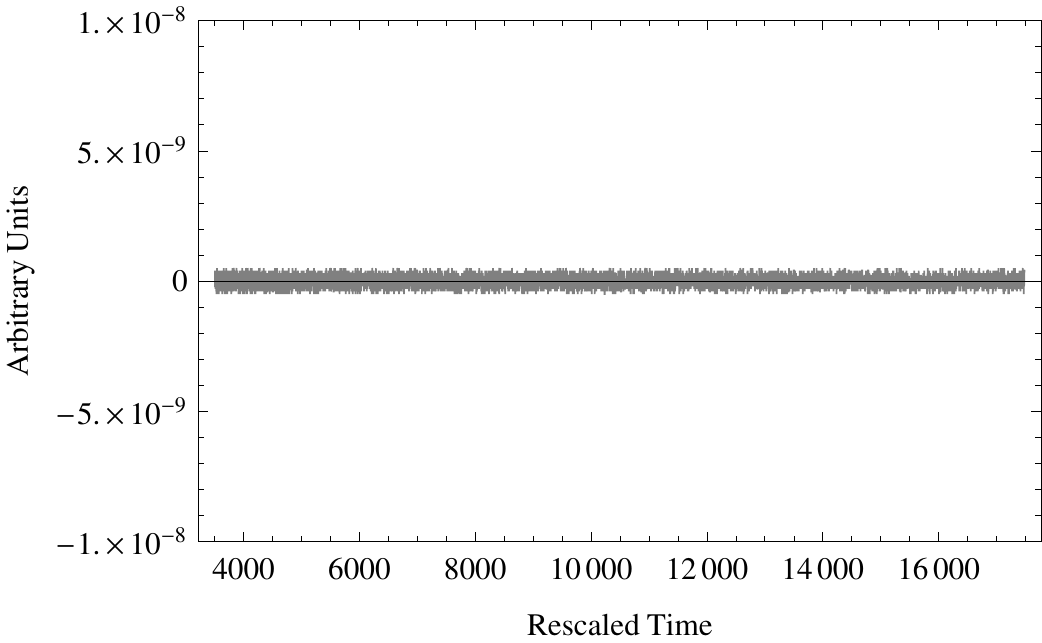}}
\caption{The upper plot shows the magnitude, in arbitrary units, versus rescaled time of the diagonal components, $h_{11}$ (solid), $h_{12}$ (dotted) and $h_{13}$ (dashed) for a mode corresponding to $|k|= 1.4 \times 10^8$ today for a $\lambda \phi^4$ model.  The lower plot shows $k_1h_{11}(k)+k_2h_{12}(k)+k_3h_{13}(k)$, which demonstrates that  perturbation is transverse via (\ref{ttcondition}).  \label{transverse}}
\end{figure}

\begin{figure}[tbp]
\resizebox{8.4cm}{!}{\includegraphics{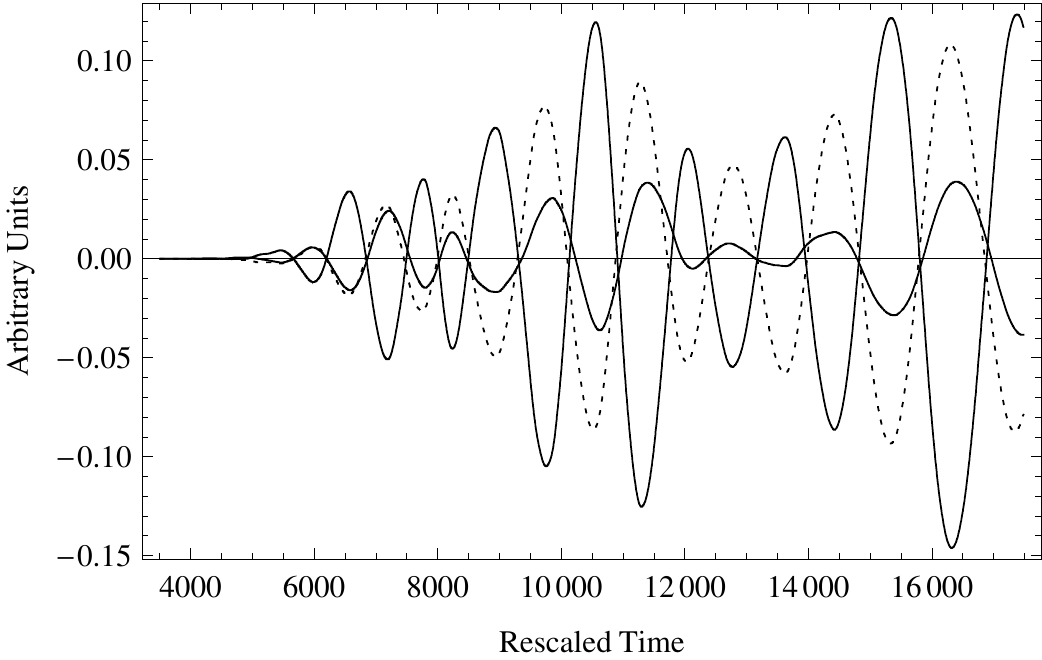}}
\resizebox{8.4cm}{!}{\includegraphics{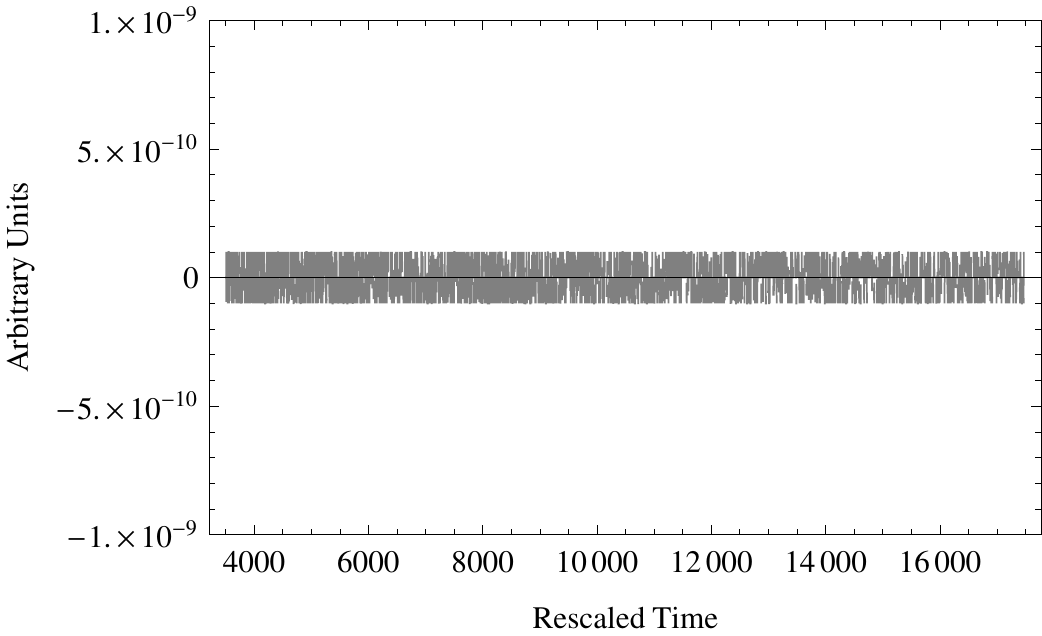}}
\caption{The upper plot shows the magnitude, in arbitrary units, versus rescaled time of the diagonal components, $h_{11}$ (solid), $h_{22}$ (dotted) and $h_{33}$ (dashed) for a mode corresponding to $|k|= 1.4 \times 10^8$ today for a $\lambda \phi^4$ model.  The lower plot shows their sum.\label{traceless}}
\end{figure}

In Figure \ref{source} we summarize the evolution of the gravitational wave background for a $V(\phi) = \lambda \phi^4/4$ model, which is the {\em de facto\/} testbed for this type of calculation, even though the underlying inflationary model is not consistent with recent CMB data. The specific model we solve is given by equation (\ref{quartic1}), with $\lambda = 10^{-14}$ and $g^2/\lambda = 120$.  We explicitly check that the numerical evolution of $h_{ij}$ respects the transverse-traceless condition. Figures \ref{transverse} and \ref{traceless} explicity show representative Fourier  components of the metric perturbation, along with explicit plots showing that that the numerical evolution preserves the transversivity and tracelessness of $h_{ij}$.   Note that the transversivity  condition (\ref{ttcondition}) takes the form $k_1h_{11}(k)+k_2h_{12}(k)+k_3h_{13}(k)$ in momentum-space.

\begin{figure}[tbp]
\resizebox{8.4cm}{!}{\includegraphics{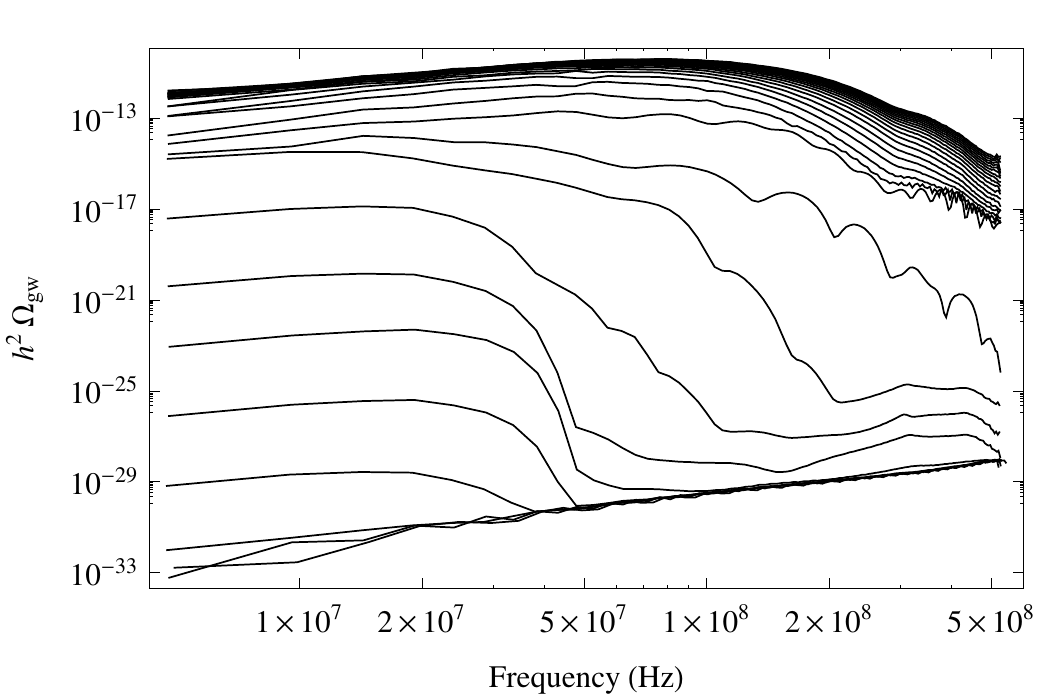}}
\caption{The evolution of the gravitational wave spectrum for a quartic inflation model, where $\lambda = 10^{-14}$ and $g^2/\lambda = 120$.  \label{timeshots}}
\end{figure}

\begin{figure}[tbp]
\resizebox{8.4cm}{!}{\includegraphics{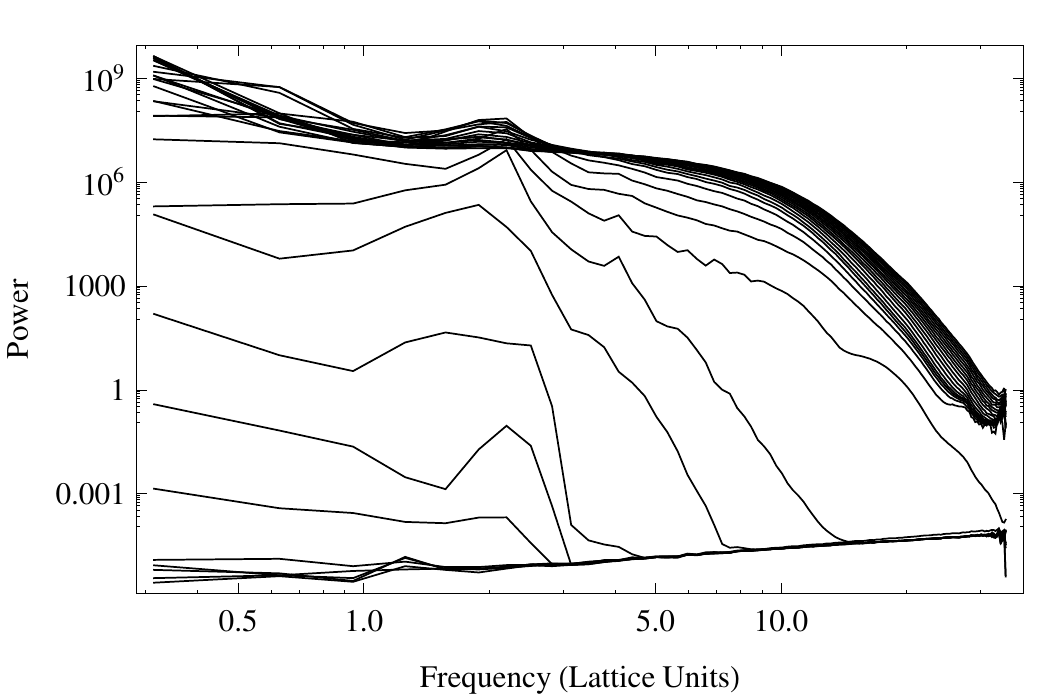}}
\resizebox{8.4cm}{!}{\includegraphics{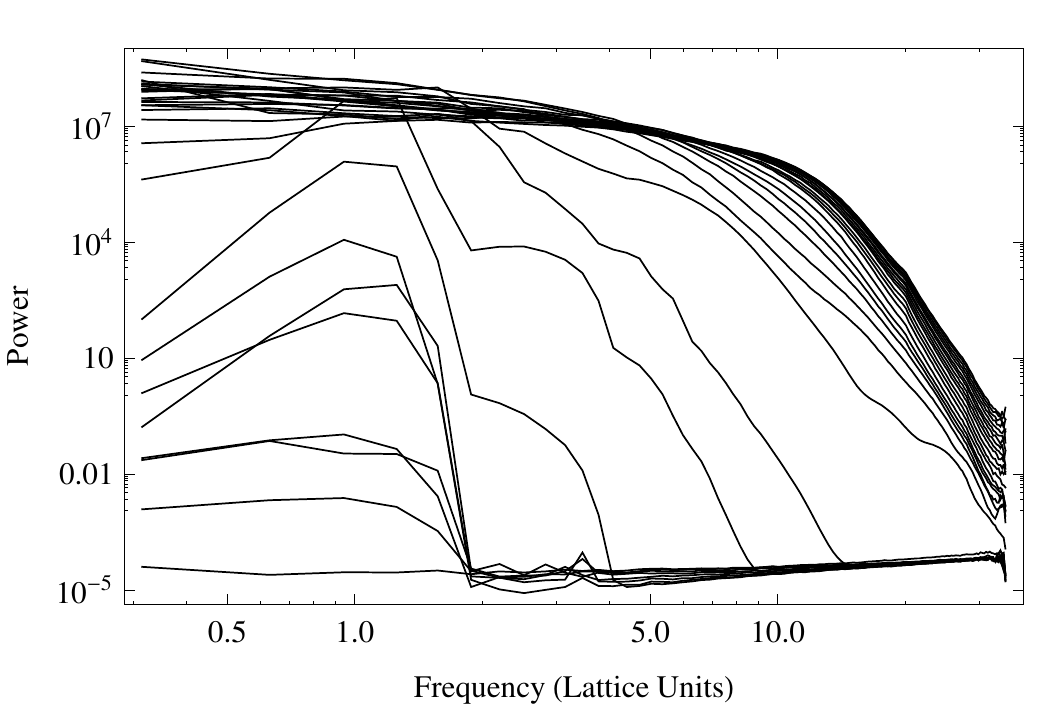}}
\caption{The evolution of the power spectra of the fields for both $\phi$
  (above) and $\chi$ (below).  This is done for the case where $\lambda =
  10^{-14}$ and $g^2/\lambda = 120$ on a $128^3$ lattice.\label{fieldspectra}}
\end{figure}

Unlike \cite{Khlebnikov:1997di} and \cite{Easther:2006gt}, we can compute the gravitational wave spectrum at any point during the simulation -- a feature shared by the leapfrog \cite{GarciaBellido:2007dg} and Green's function algorithms \cite{Dufaux:2007pt}.    Figure \ref{timeshots} shows the evolution of the gravitational wave spectrum for $V(\phi) = \lambda \phi^4$  model with $\lambda = 10^{-14}$ and $g^2/\lambda = 120$.    The final amplitude of this spectrum is noticeably smaller than that computed using the ``box" algorithm  \cite{Easther:2006gt}. However, this algorithm did not fully incorporate the expansion of space in its computation of the gravitational wave spectrum, and the universe undergoes significant expansion during the resonant phase, and this expansion will have the effect of diluting the gravitational waves, and reducing their final amplitude.    One can clearly see the ``pumping'' of the low frequency modes as resonance begins, and then the growth of the higher frequency modes as it continues.  The 
evolution of $h_{ij}$  is  driven only by the inhomogeneities in the fields, and is not directly sourced by the potential. The overall form of the field's power spectra, shown in Figure \ref{fieldspectra},  closely resembles that of the gravitational waves.

Our procedure applies to any source of stochastic gravitational radiation where the signal is generated by large scale inhomogeneities, such  as cosmological phase transitions \cite{Kosowsky:1992rz} and more specifically large scale bubble collisions \cite{Kosowsky:1991ua}. Thus provided these processes can be simulated numerically, this algorithm could be adapted to compute the gravitational radiation they generate. For the latter case, the crucial limitation is the resolution of the lattice itself: to accurately model the process of bubble nucleation till bubble collision, the distance between two neighbouring lattice grid points must be smaller than the nucleation size of the bubble, while the total lattice size must be larger than the bubble radius at percolation.  

\subsection{Experimental Prospects}

Most theoretical discussions of gravitational wave spectra are expressed in terms of the ratio between the  spectral energy density of the gravitational wave and the total energy density, as measured at the present day:
\begin{equation}
\Omega_{gw}(f) = \frac{1}{\rho_c} \frac{d\rho_{gw}}{d\ln f}.
\end{equation}
Experimental bounds are commonly expressed in terms of the {\it
  spectral density}, $S_h(f)$, which is an ensemble average
  over the Fourier amplitudes of the plane-wave solutions
  $\tilde{h}_A(f,\hat{\Omega})$ averaged over all directions \cite{Maggiore:1999vm}.
  That is, a transverse-traceless metric perturbation can be decomposed into
  plane wave solutions, each carrying two polarizations, in all directions, i.e
\begin{equation}
h_{ij} = \sum_{A = +, \times}\int_{-\infty}^\infty df \int d\hat{\Omega}
\tilde{h}_A(f,\hat{\Omega})e^{-2\pi i ft }e^A_{ij}(\hat{\Omega})
\end{equation}
Here, $e^A_ij$ is a tensor that identifies the polarizations relative to the
coordinates $x^a$ and $x^b$ and $\hat{\Omega}$ is a directional unit vector
with differential, $d\hat{\Omega}=d\cos\theta d\phi$.  From this, 
\begin{equation}
\delta(f-f^\prime)\delta_{AA^\prime}\frac{1}{2}S_h(f) = \int
d\hat{\Omega}d\hat{\Omega}^\prime
\left<\tilde{h}^*_A(f,\hat{\Omega})\tilde{h}_{A^\prime}(f^\prime,\hat{\Omega}^\prime)\right>.
\end{equation}
More usefully, this quantity can be directly related to $\Omega_{gw}$ through
\begin{equation}
\Omega_{gw} = \frac{4\pi^2}{3H_0^2}f^3S_h(f).
\end{equation}
We will typically find signals where $\Omega_{gw}(f)$ is, at most, ${\cal O}(10^{-11})$. Consequently, in order to detect this signal we need a an observatory that can detect spectral density on the order of $S_h\sim 10^{-53} h^2\,{\rm Hz}^{-1}$ at $100\, {\rm Hz}$ or $S_h\sim 10^{-41} h^2\,{\rm Hz}^{-1}$ at $10^{-2}\, {\rm Hz}$.  For comparison, the minimum spectral density detectable by Advaced LIGO is approximately $10^{-48}\, {\rm Hz}^{-1}$ at $100 \,{\rm Hz}$ \cite{ligopsd}, and LISA is $10^{-40} \,{\rm Hz}^{-1}$ at $10^{-2}\, {\rm Hz}$ \cite{lisapsd}.

\section{Cosmological Models}

\subsection{Quadratic Inflation}

\begin{figure}[htbp]
\resizebox{8.2cm}{!}{\includegraphics{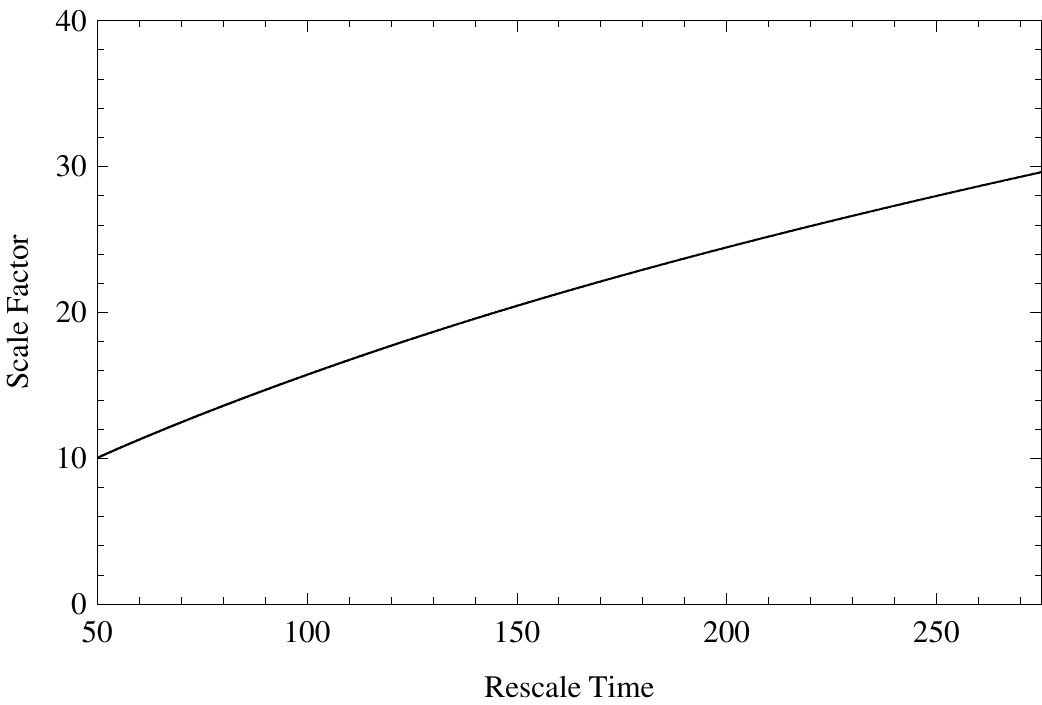}}
\resizebox{8.2cm}{!}{\includegraphics{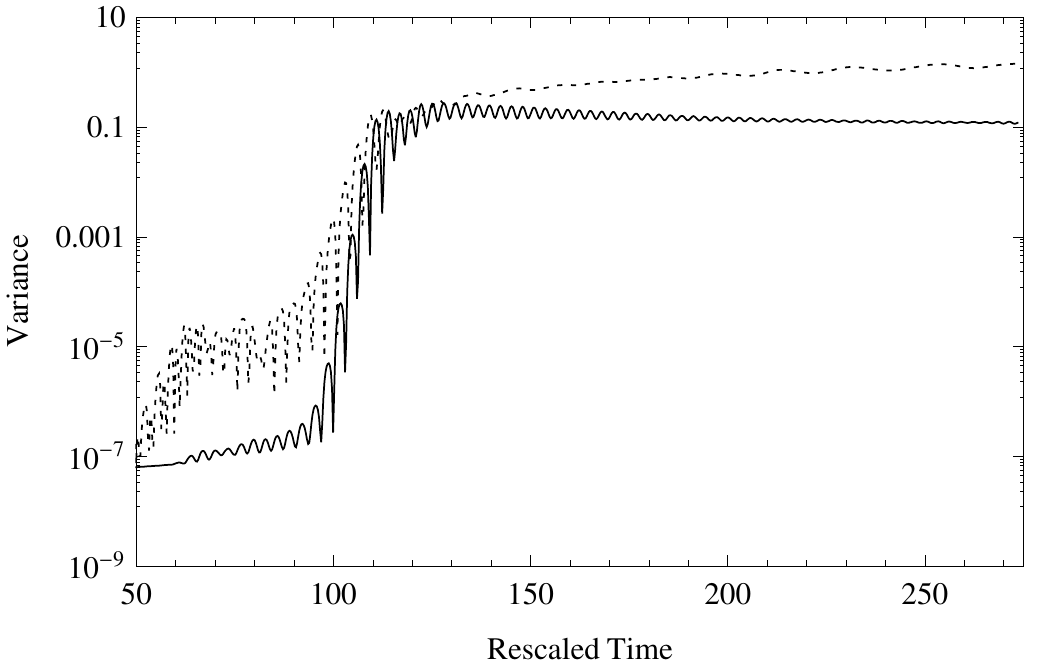}}
\resizebox{8.2cm}{!}{\includegraphics{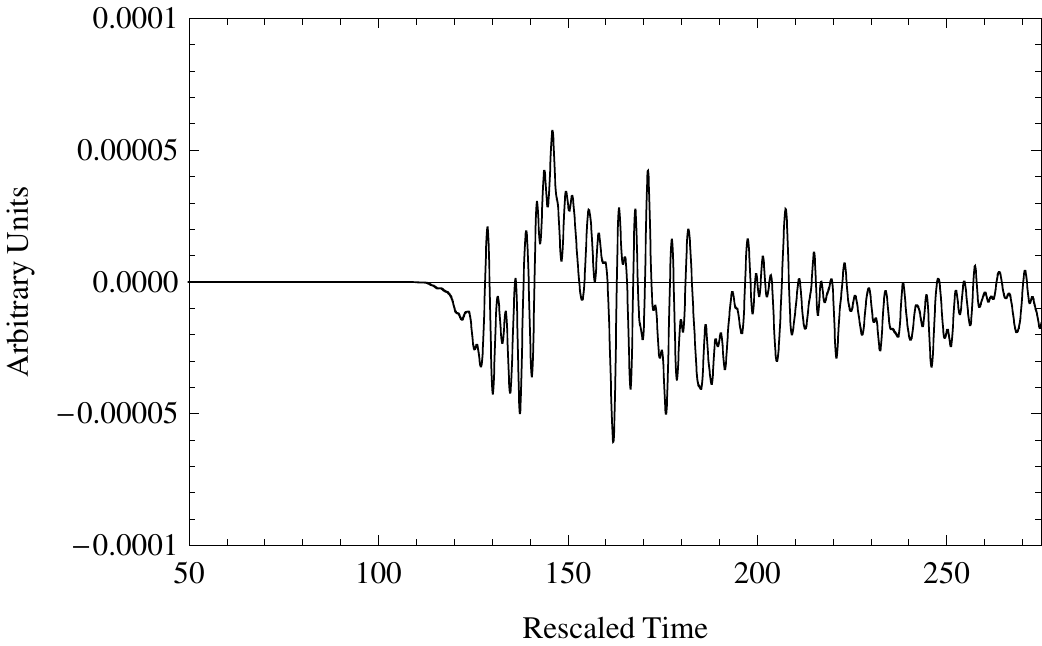}}
\resizebox{8.2cm}{!}{\includegraphics{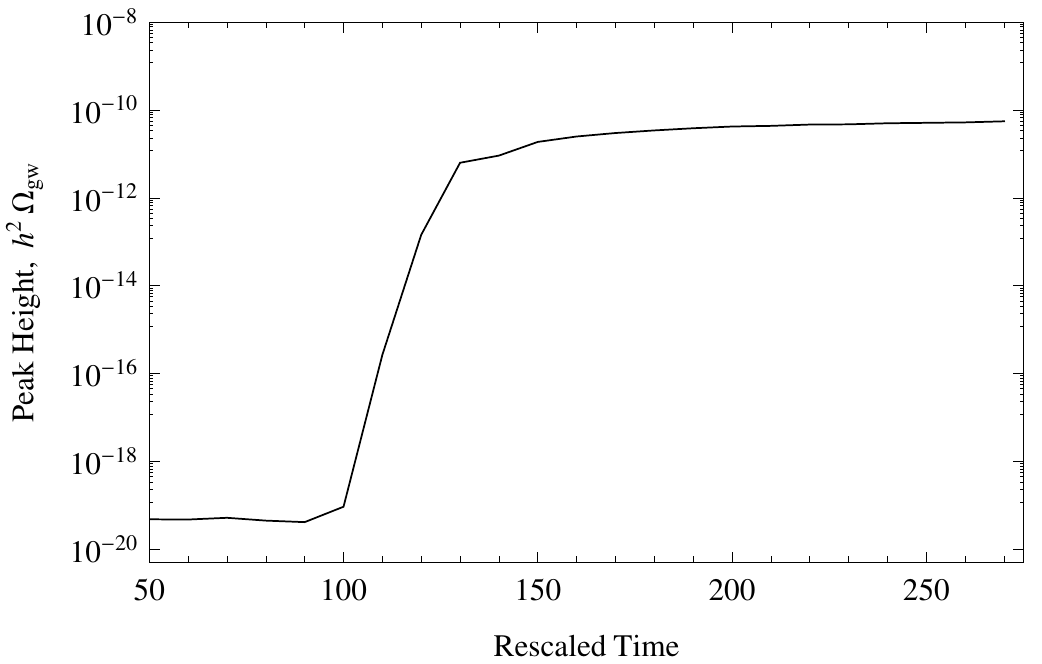}}
\caption{From top to bottom: The scale factor, the variances of $\phi$ (dotted) and $\chi$
  (solid), $S_{11}^{TT}$ for a mode corresponding to $|k|=  5.4\times 10^8\,{\rm Hz}$ today, and the maximum height of the gravitational wave spectrum as a function of time. This is for the $m = 10^{-6}m_p$ and $g^2m_{pl}^2/m^2 = 2.5\times10^5$ model.\label{m2}}
\end{figure}

Having established our overall methodology, we now compute representative spectra for several explicit cosmological models.  In Section \ref{evofpert}, we presented results for a chaotic model with a quartic potential, so as to best compare our calculations to those of other groups.  We now consider the potential 
\begin{equation}
V(\phi) = \frac{1}{2}m^2 \phi^2
\end{equation}
which was also treated in \cite{Easther:2006gt,Easther:2006vd}.   The overall evolution of this system is summarized in Figure~\ref{m2}, while the spectrum is similar to that obtained using the box algorithm \cite{Easther:2006gt}.  In this case the agreement between the box algorithm and the spectral code is considerably better, and we see that the box code actually underestimates the maximal power in gravitational waves.  Note that the total growth of the universe during resonance is smaller in this case than it was during quartic inflation -- and since the box algorithm effectively compute $h_{ij}$ in flat space, the discrepancy introduced by this approximation is accordingly reduced.   

\subsection{Low-Scale Inflation}

In inflationary models with a single free parameter in the potential, the scale of inflation is typically fixed by matching to the observed amplitude of the scalar perturbation spectrum. Consequently, for the quadratic potential we discussed above, $m$ is not a free parameter.     Now consider  a general hybrid inflation model \cite{Linde:1993cn,Copeland:1994vg,GarciaBellido:1998gm} ,
\begin{equation}
V (\phi, \sigma) = \frac{(M^2-\lambda\sigma^2)^2}{4\lambda} +
\frac{m^2}{2}\phi^2 + \frac{h^2}{2} \phi^2\sigma^2 +  \frac{g^2}{2} \phi^2\chi^2  .
\end{equation}
So long as $\phi$ is large, $\sigma=0$ is a stable minimum and the energy
density is dominated by the potential energy associated with $\phi$.  
When $\phi \approx M/h$, $\sigma=0$ is no longer a minimum and $\sigma$ quickly settles
into one of the two new minima,  $\sigma = \pm M/\sqrt{\lambda}$ a process known as the ``waterfall'' phase transition. We assume that the field $\sigma$  settles coherently into its
vacuum expectation value and there is no $\sigma$ particle production.  This
is consistent with the assumption that $m \ll M$. 
Assuming, with no loss of generality, that the field settles into its positive
minimum, $\phi$ acquires the following effective potential energy,  
\begin{equation} \label{HVeff}
V(\phi) = \frac{1}{2}\left(m^2+\frac{h^2M^2}{\lambda}\right)\phi^2,
\end{equation}
or, since we have assumed a small $m$,
\begin{equation}
V(\phi) = \frac{1}{2}\left(\frac{h^2M^2}{\lambda}\right)\phi^2.
\end{equation}
In this regime, we recover a chaotic inflation model where $\phi$ acquires an {\it effective}
mass, $m_{\mbox {eff}} = h^2 M^2/\lambda$. This system was studied in detail in  \cite{Easther:2006vd} by the present authors.  In particular, we used this model to demonstrate that the spectrum of gravitational radiation produced during resonance has the scaling properties conjectured in \cite{Easther:2006gt}.  However, this model has the disadvantage of needing exceptionally small parameter values in order to sure that the structure of resonance does not depend directly on the inflationary scale.  

\subsection{General Hybrid Potentials}

Now consider the potential
\begin{equation}
V(\phi,\chi) = 
\frac{\lambda_\phi}{4}\phi^4+\frac{\lambda_\chi}{4}\chi^4 + \frac{g}{2}
\phi^2\chi^2,
\end{equation}  
where we allow $g$ to aquire either sign.   In principle  \cite{Greene:1997ge}, one can include mass terms for both the $\phi$ and the $\chi$ fields. We omit them here for simplicity, and we  one must also ensure that if $g<0$, $V(\phi,\chi) \ge 0$ everywhere, or
\begin{equation}
\frac{\lambda_\phi \lambda_\chi}{g^2} > 1.
\end{equation}
When $\phi$ is large the $\chi$ field has a substantial effective mass and remains at $\chi = 0$.  If  $g>0$, the effective mass of $\chi$ only vanishes at the origin of the potential. This suppresses the production of $\chi$ particles and leads to the slow and inefficient production of $\phi$ particles.  Conversely,  for $g<0$ the $\chi$ field picks up a tachyonic mass when $\phi^2$ is small, there are rapid oscillations in both  $\phi$ and $\chi$, and particle production proceeds at a dramatic pace.  While  Figure \ref{stabilitychart} strictly applies to the Mathieu equation, this system effectively has a  negative $q$ and the resonant modes thus live in the lower portion of the plot, where the instability regions are broad, and the imaginary part of the critical exponent is large. Consequently, resonance occurs extremely promptly, and the resulting dynamics differ dramatically from  those of all the other models considered here.  In particular, the universe grows by a factor of $\sim3$ as the vast majority of the gravitational wave power is generated. Further, this model produces $\phi$ quanta, and the corresponding variance has a sharp peak, rather than the broad ``hump'' seen in the other cases we examine.  

 \begin{figure}[htbp]
\resizebox{8.4cm}{!}{\includegraphics{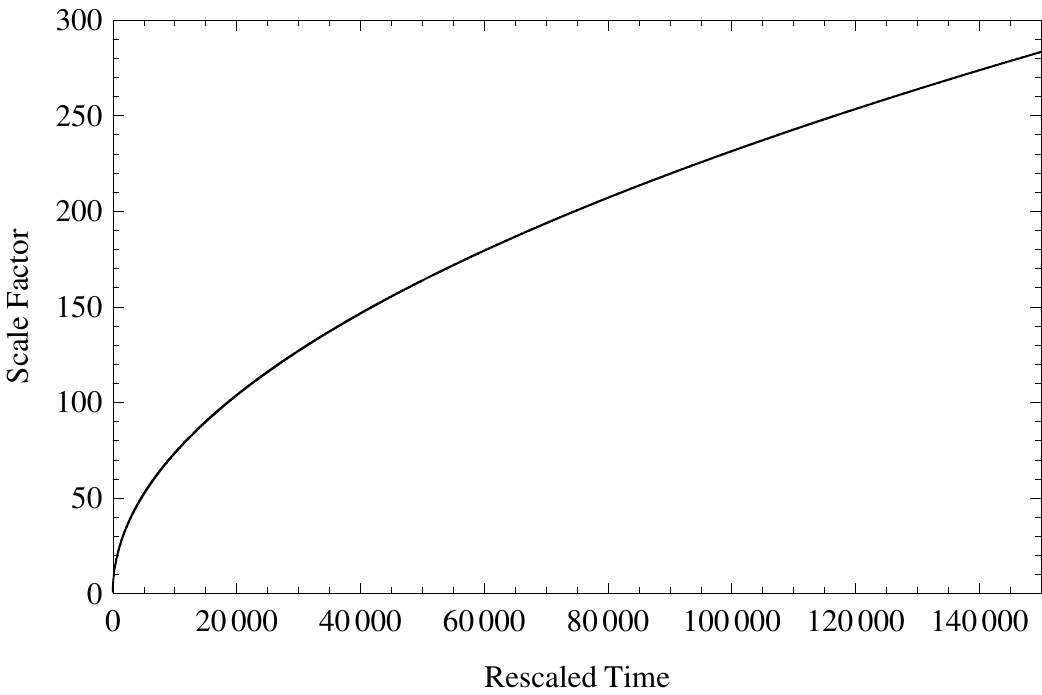}}
\resizebox{8.4cm}{!}{\includegraphics{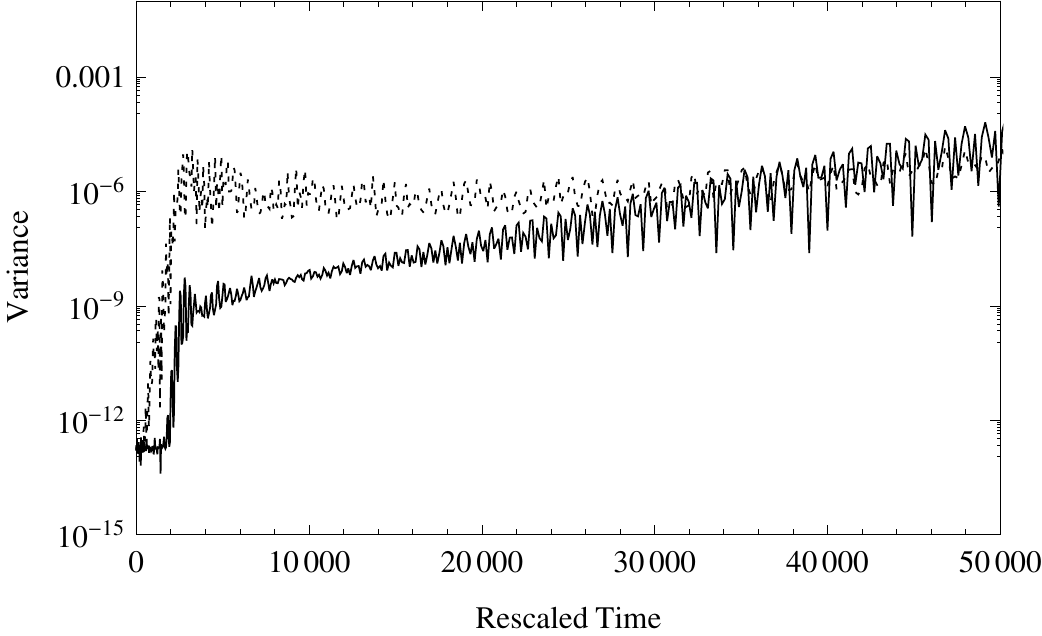}}
\resizebox{8.4cm}{!}{\includegraphics{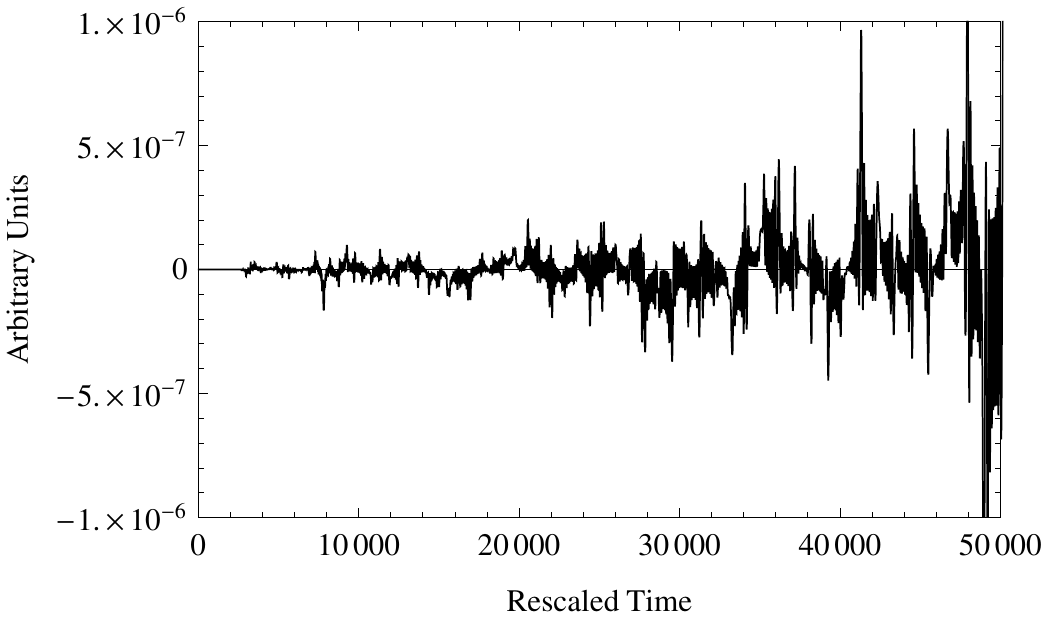}}
\resizebox{8.4cm}{!}{\includegraphics{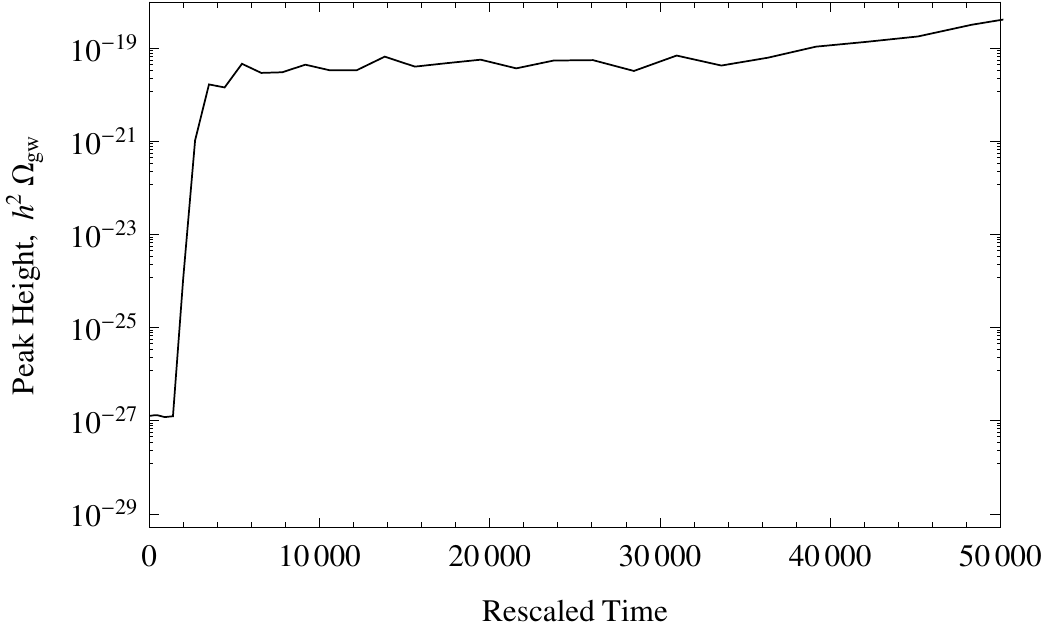}}
\caption{From top to bottom: The scale factor (normalized to one at the
  beginning of the simluation), the variances of $\phi$ (solid) and $\chi$
  (dotted), and one component, $S_{11}^{TT}$ for a mode corresponding to $|k|\approx5.4\times 10^7 \,{\rm Hz}$ today, and the maximum height of the gravitational wave spectrum as a function of time.  This is for the $g =10^{-10} $ case.  \label{pos}}
\end{figure}\begin{figure}[htbp]
\resizebox{8.4cm}{!}{\includegraphics{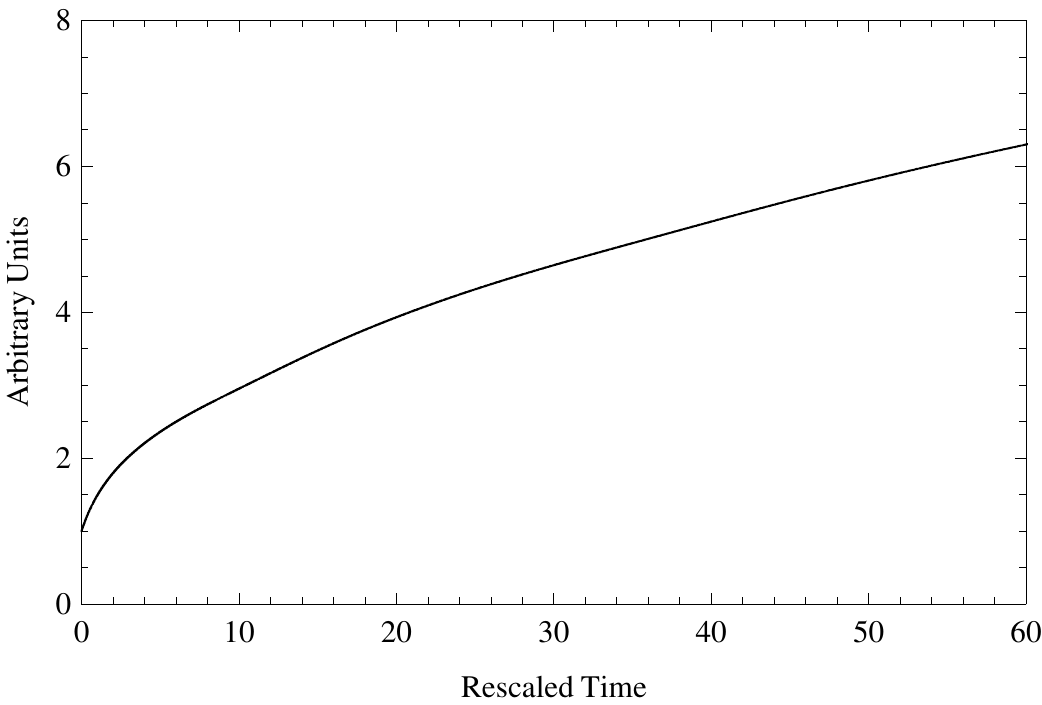}}
\resizebox{8.4cm}{!}{\includegraphics{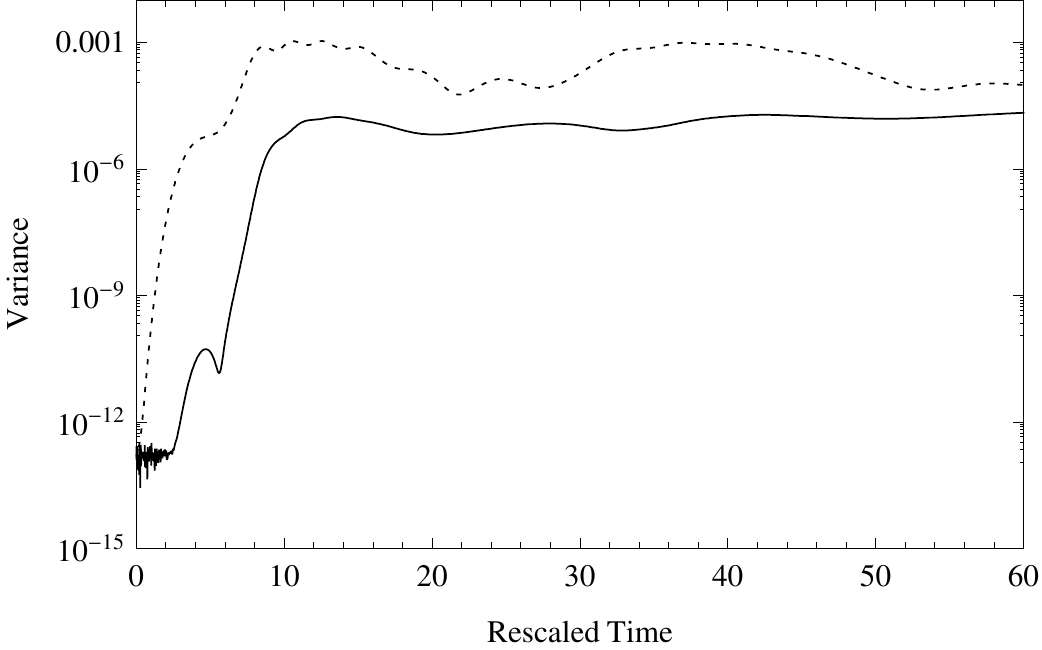}}
\resizebox{8.4cm}{!}{\includegraphics{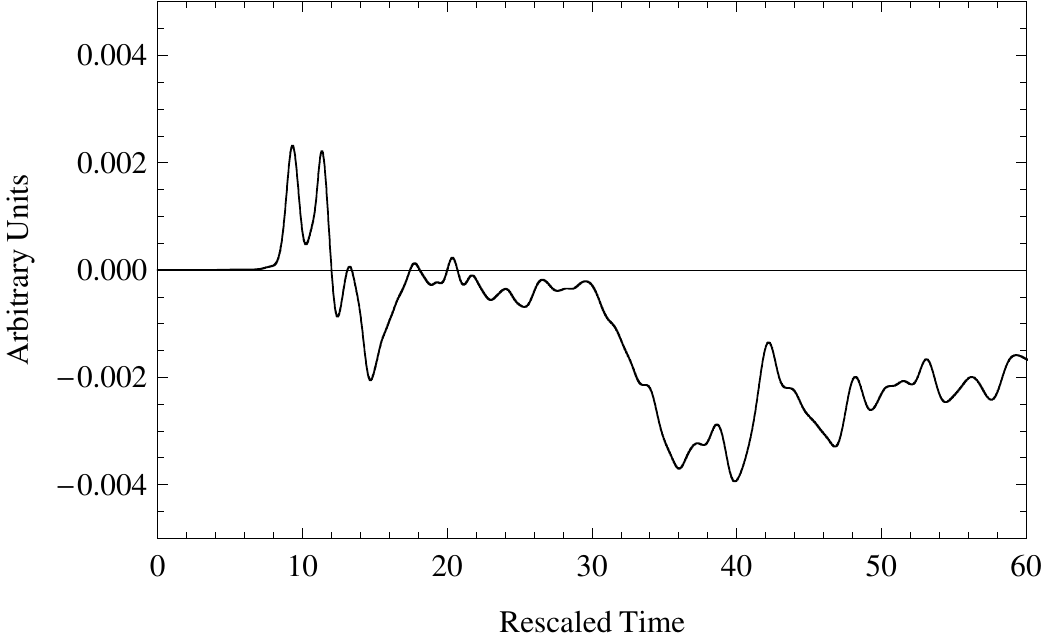}}
\resizebox{8.4cm}{!}{\includegraphics{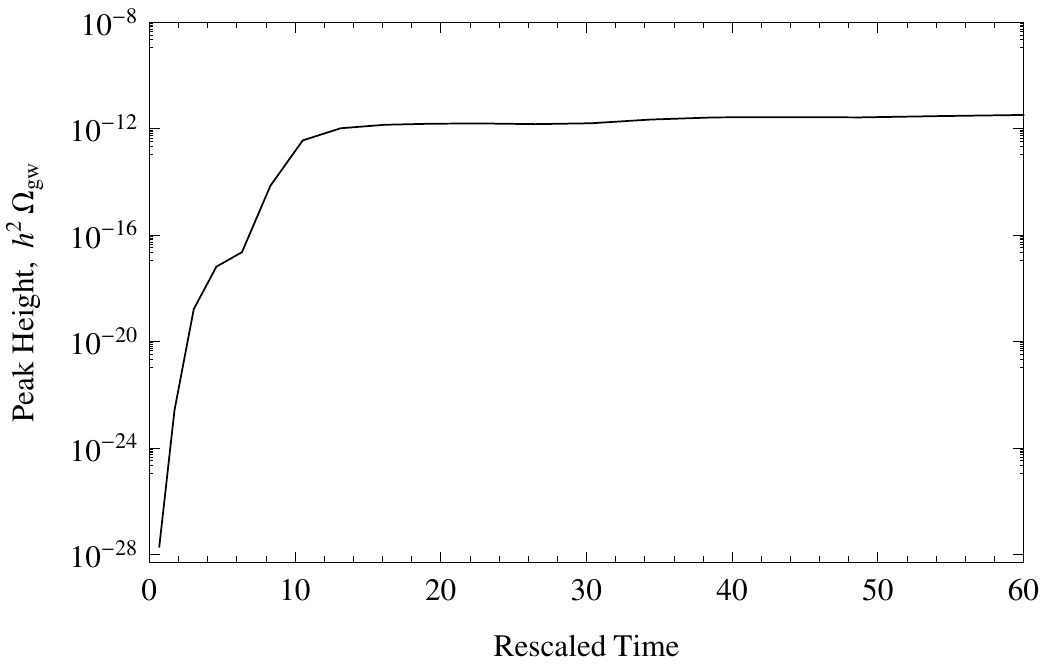}}
\caption{From top to bottom: The scale factor (normalized to one at the
  beginning of the simluation), the variances of $\phi$ (solid) and $\chi$
  (dotted), and one component, $S_{11}^{TT}$ for a mode corresponding to $|k|\approx 5.2 \times 10^7 \,{\rm Hz}$ today, and the maximum height of the gravitational wave spectrum as a function of time.  This is for the $g=-10^{-10}$ case. \label{neg}}
\end{figure}

We have chosen to work with the same parameters used in the simulations of  \cite{Greene:1997ge}, $\lambda_\phi = 10^{-12}$, $\lambda_\chi = 10^{-7}$, and $\left|g\right| = 10^{-10}$, which provides a cross-check on our evaluation of the field evolution.   We summarize the resonant epochs in these models in Figures \ref{pos} and \ref{neg}, for positive and negative $g$ respectively.  The variances in Figure \ref{pos} and Figure \ref{neg} are consistent with Figures 4 and 5 of \cite{Greene:1997ge}.   With a negative coupling constant the fields quickly become inhomogeneous. Conversely, with a positive coupling case the $\chi$ field initially becomes inhomogeneous, with the $\phi$ following considerably later, leading to two superimposed sets of gravitational waves, as can be seen from the ``crossing'' of the two variances  in Figure~\ref{pos}.

\begin{figure}[tbp]
\resizebox{8cm}{!}{\includegraphics{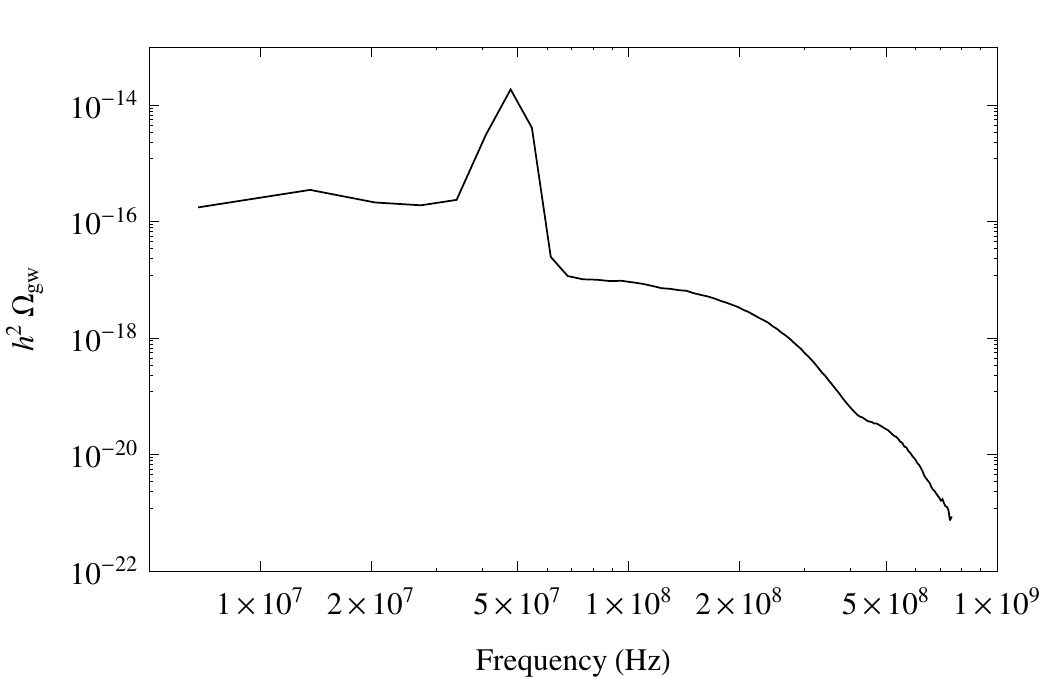}}
\resizebox{8cm}{!}{\includegraphics{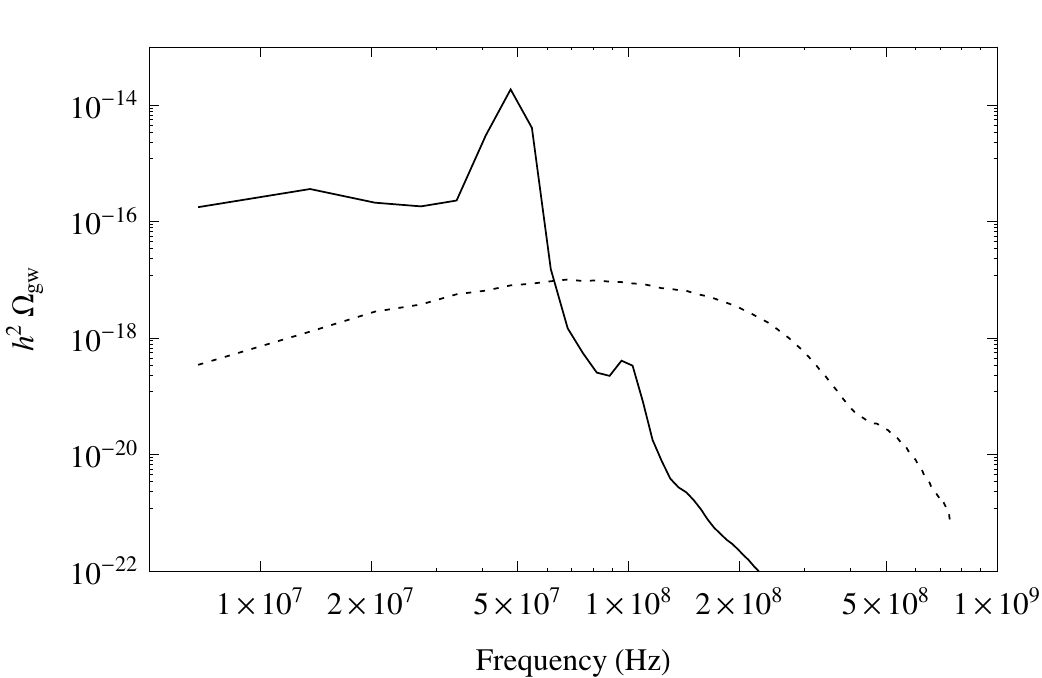}}
\caption{The overall spectrum (top panel) and the individual spectra (bottom panel) for the simulation shown in Figure~\ref{pos}.  In the lower plot, the dotted line shows  gravitational
waves due to $\phi$-particle production while the solid line denotes the contribution from the $\chi$ field. \label{poscasespec}}
\end{figure}
 \begin{figure}[tbp]
\resizebox{8cm}{!}{\includegraphics{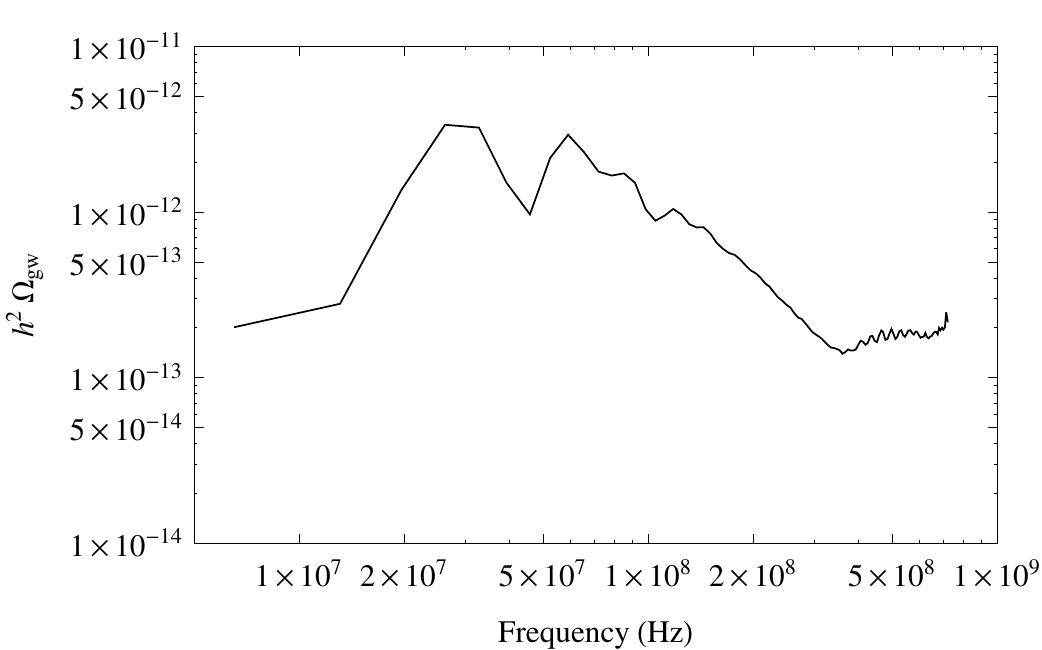}}
\caption{The spectrum of gravitational waves generated during preheating with a negative coupling, with the same parameters as the simulation shown in Figure~\ref{neg}. \label{negcasespec} }
\end{figure}

We display the gravitational power spectra of these models in Figures \ref{poscasespec} and
\ref{negcasespec}. In the positive coupling case, we separate the contribution from the two fields, and show that these sum together to give a spectrum with a pronounced peak. Conversely, when the coupling is negative, particle production occurs very rapidly, and the associated gravitational wave spectrum again has a substantial amplitude, despite the very different resonance dynamics.  We plan to examine this system more carefully in a future publication.

\section{Numerical Issues}

The source term for $h_{ij}$ contains differences of combinations of derivatives which are evaluated on a discretized lattice, there is ample opportunity for numerical ``noise'' to contaminate these simulations.   Consequently, we have to take care that the spatial extent of our lattice (henceforth the ``box'') is small enough  to ensure that highest Fourier modes that contribute to the gravitational wave spectrum are well resolved.

\begin{figure}[tbp]
\resizebox{8cm}{!}{\includegraphics{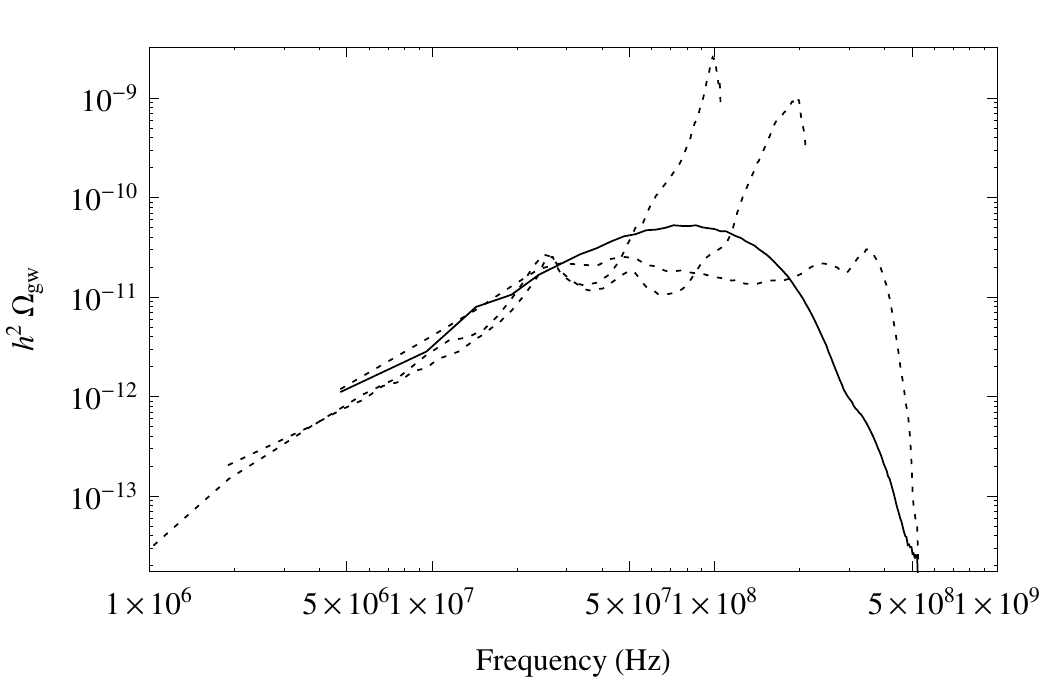}}
\caption{This shows a test of the integration step size.  The dotted lines show simulations, with varying box sizes, where the metric perturbations are evolved at every timestep, the solid line shows the case where the timestep for the metric perturbation is increased.   \label{noise}}
\end{figure}

 In Figure~\ref{noise} we show the consequences of making the box too large, and thus losing the fine resolution needed to properly resolve the peak. The specific model here is $V(\phi)=\lambda\phi^4/4$ with $g^2/\lambda=120$ as in Section \ref{gravpow}.   We see that the discretization noise typically adds spurious power to the spectrum. This issue does not appear to explain the discrepancy between the box algorithm and the spectral algorithm discussed in Section \ref{gravpow} -- the lattice used in \cite{Easther:2006gt} is large enough to resolve the peak, and the ``box'' algorithm yields a much better match to the results found for a quadratic potential.   However, it is noteworthy that the universe grows significantly during the reheating phase following inflation with a quartic potential, and this growth would tend to dilute the gravitational waves as they are produced -- and the box method does not properly account for the growth of the universe in its computational of the gravitational waves. Consequently, we conclude that it is safest to solve the $h_{ij}$ self-consistently in an expanding background when evaluating the gravitational wave signal generated during preheating. Further the computational cost of doing so is relatively small.

\begin{figure*}[t]
\resizebox{15cm}{!}{\includegraphics{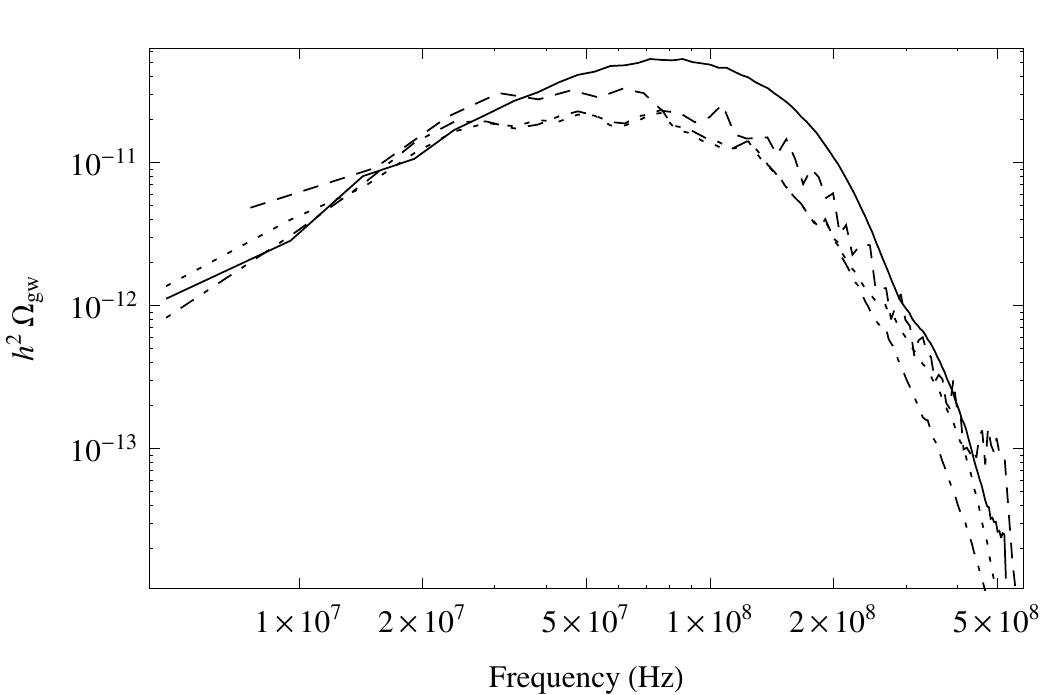}}
\caption{We show the results of simulation a quartic inflation with $g^2/\lambda=120$ using both our spectral code, and the Green's function algorithm. This simulation matches Figure 4 of  \cite{Dufaux:2007pt}, and the Green's function results are shown with a dotted line. We show results from the spectral code for three different lattice sizes -- $128^3$ (solid), $256^3$ (dashed) and $512^3$ (dot-dashed). The last two lines overlap, so we are presumably approaching the continuum limit, and we also have excellent agreement with the results of   \cite{Dufaux:2007pt}, in which the gravitational wave spectrum was obtained using a very different numerical algorithm.   \label{compare}}
\end{figure*}

In Figure~\ref{compare} we compare results from our spectral code to those obtained using the Green's function algorithm of Dufaux {\em et al.\/} \cite{Dufaux:2007pt}, and find that the two methods overlap exceptionally well.   For the ``standard'' $128^3$ lattice used in the calculations in this paper, the agreement between the two methods is good but not perfect. We also present data from $256^3$ and $512^3$ runs which agree very closely with one another, and are presumably close to the continuum limit.\footnote{Hal Finkel collaborated with us on writing the code used for the $256^3$  and $512^3$ runs, and we thank Gary Felder with providing us with the raw data used to plot the Green's function spectrum in Figure~\ref{compare}.}   Moreover, these results also overlap closely with the Green's function result, which provides a crucial cross-check oft the algorithms used by ourselves and the authors of   \cite{Dufaux:2007pt}.  Note that both codes use the same transfer function to shift these spectra into their presently observable values.      Moreover, the results of \cite{GarciaBellido:2007af} appear to be in good qualitative agreement with those presented here.

The numerical simulations here are conceptually straightforward -- we are, after all, simply evolving a set of second order partial differential equations, but experience has shown that their actual implementation can require some finesse. We now have solid agreement between two independent codes -- the discrepancies between the two {\em algorithms} shown in Figure~\ref{compare} are of the same order as the changes induced in the spectrum by modifying the size of the underlying lattice. One can thus be confident that the gravitational wave signal generated during preheating is being accurately evaluated, and these results can be safely used to assess observational strategies for detecting a stochastic background.   

We do note, however,  that almost all the different numerical approaches summarized in the introduction have used a staggered leapfrog algorithm to evolve the {\em fields\/}, and it would be worth exploring the consequences of using different algorithms for this part of the calculation, particularly those with  better than second order accuracy in their spatial differencing.  In particular, since we are using a spectral algorithm for the metric perturbations, it is a natural extension of our approach to  implement a spectral solver for the field evolution as well. This is somewhat complicated since the couplings between the fields render these equations nonlinear, but the application of spectral methods to nonlinear systems is a well understood problem.    

As presently implemented, our code uses a good deal of memory -- each field requires two ``copies'' of the grid (the field, and its first time derivative), plus 12   more to evolve the Fourier modes of the  6 $h_{ij}$, and a further handful for computing source terms and Fourier transforms. If we assume isotropy (which is a very reasonable expectation), equation~\ref{omega} is a sum over three equivalent diagonal terms and three equivalent off-diagonal terms.  Consequently, we can reduce the labor and storage involved in computing the $h_{ij}$ by a factor of three if we assume that $h_{11} = h_{22} = h_{33}$ and $h_{12}=h_{13}=h_{23}$, and this works as expected in practice.   We could further improve matters by only evolving a subset of the Fourier modes of the $h_{ij}$ (a strategy employed by  \cite{Dufaux:2007pt}) -- the only downside to these approaches is that, if carried too far, the spectrum acquires significant stochastic noise, since it is being computed from a small sample of total set of modes.  

The evolution code of the fields will  have some characteristic timestep $dt$, whose optimum value is a function of the specific dynamical system, and the lattice spacing $dx$.  Conversely, most of the modes of the $h_{ij}$ can be evolved with a significantly larger timestep, so one has the option of updating the $h_{ij}$ evolution on a different timescale to that of the fields, further improving performance.\footnote{This strategy is particularly suited to cases where one evolves the Fourier modes of the $h_{ij}$ since the mode equations are independent at linear order in perturbation theory.  If one chooses to solve $\bo h_{ij}$ on a spatial grid the timestep will need to be commensurate with the lattice spacing in order to satisfy the Courant condition, unless one ``downsamples'' the grid for  the metric terms, relative to that of the fields.}   In this analysis, which is essentially a ``proof of concept'', we have not aggressively pursued these optimizations but we do plan to employ them in our future work.  In our current implementation, the evaluation of $h_{ij}$ takes approximately 90\% of the overall runtime, so any optimization to this calculation will have substantial benefits.

\section{Discussion}

This paper describes the {\em spectral algorithm\/} for computing a stochastic background of gravitational waves generated during an epoch in which the universe is highly inhomogeneous on small scales.  We have focussed on the spectrum generated at the end of inflation, during an era of preheating and/or parametric resonance. However,  this  algorithm is applicable to any epoch in which  inhomogeneities source gravitational radiation, including first order phase transitions \cite{Kosowsky:1992rz}, bubble collisions \cite{Kosowsky:1991ua,Caprini:2007xq}, or  MHD turbulence \cite{Kosowsky:2001xp,Kahniashvili:2005qi}.    

We began this paper with a review of other recent approaches to this problem, highlighting their relative merits.  In addition to the previous work of the current authors \cite{Easther:2006gt,Easther:2006vd}, two other independent codes have been recently been developed to address this problem  \cite{GarciaBellido:2007dg,GarciaBellido:2007af,Dufaux:2007pt} and we explicitly compare the results of our spectral algorithm that the Green's function algorithm of \cite{Dufaux:2007pt}, showing that the difference between the computed spectra is of the same order as that introduced by the finite resolution of  the underlying spatial lattice.  This is a welcome development, as it validates both algorithms and their explicit implementations. Conversely, algorithms for evaluating the gravitational wave spectrum which do not account for the expansion of the universe do not necessarily overlap with the results obtained with our spectral algorithm, and should not be used if the universe expands by a large (${\cal O}(10)$ or more) factor during the resonant phase.
 
The overall scaling properties of any stochastic spectrum produced as inflation ends and the universe reheats were first conjectured in  \cite{Easther:2006gt}. These claims were confirmed explicitly for a simple resonant model in \cite{Easther:2006vd}, using the algorithm we describe here.    In this paper we extend our analysis to a class of hybrid models and  find spectra that are qualitatively different from those seen previous in this context. In particular, models with an ``inverted'' coupling, first discussed in \cite{Greene:1997ge}, have a particularly strong period of resonance. This is a significant development for several reasons -- firstly, resonance here is substantially different from that of the chaotic models investigated previously. Secondly, this model provides  a more  realistic example of resonance with an arbitrary inflationary scale. We plan to investigate its properties  more carefully in a future paper, in order to better understand the correlation between they dynamics of resonance and preheating and the resulting gravitational wave background.   

In this paper, for the sake of computational convenience most of our calculations were performed on $128^3$ grids, which easily fits on a single computational node. However, our spectral code has been extended to a cluster environment, and which allows simulations on much larger grids.  We plan to use this code to investigate the full dynamic range of the spectrum, especially in models where the post-inflationary  Hubble scale is not vastly larger than the scales which undergo resonance.  Finally, we also plan to apply our algorithm to other situations where significant gravitational wave backgrounds can be generated by small-scale inhomogeneities in the primordial universe.
 
\section*{Acknowledgments}
We are particularly indebted to Gary Felder for a number of useful conversations, and for his and Igor Tkachev's work  on {\sc LatticeEasy\/}. We thank Hal Finkel and Doug Swanson for a number of useful conversations, and their contributions to the parallelized version of our code, which was used to generate plots shown in Figure   \ref{compare} and we are grateful to Amanda Bergman, Latham Boyle, Jean-Fran\c{c}ois Dufaux, Daniel Figueroa, Juan Garc{\'\i}a-Bellido, Lev Kofman and Jean-Philippe Uzan for a number of useful discussions.  This work is supported in part by  the United States Department of Energy, grant DE-FG02-92ER-40704.

\end{document}